\documentclass[twocolumn]{aastex631}

\usepackage{amsmath}
\usepackage{graphicx}

\newcommand{\civ}{C\,{\sc iv}}

\newcommand{\mgii}{Mg\,{\sc ii}}

\newcommand{\oiii}{[O\,{\sc iii}]}
\newcommand{\hb}{H$\beta$}
\newcommand{\hgamma}{H$\gamma$}

\shorttitle{Luminous Quasars and Their Host Galaxies at $z\gtrsim6$}
\shortauthors{Yue et al.}
\graphicspath{{./}{figures/}}

\begin{document}

\title{EIGER V. Characterizing the Host Galaxies of Luminous Quasars at $z\gtrsim6$}

\correspondingauthor{Minghao Yue}
\email{myue@mit.edu}

\author[0000-0002-5367-8021]{Minghao Yue}
\affiliation{MIT Kavli Institute for Astrophysics and Space Research, 77 Massachusetts Ave., Cambridge, MA 02139, USA}

\author[0000-0003-2895-6218]{Anna--Christina~Eilers}
\affiliation{MIT Kavli Institute for Astrophysics and Space Research, 77 Massachusetts Ave., Cambridge, MA 02139, USA}

\author[0000-0003-3769-9559]{Robert A. Simcoe}
\affiliation{MIT Kavli Institute for Astrophysics and Space Research, 77 Massachusetts Ave., Cambridge, MA 02139, USA}

\author[0000-0003-0417-385X]{Ruari Mackenzie}
\affiliation{Department of Physics, ETH Z{\"u}rich, Wolfgang-Pauli-Strasse 27, Z{\"u}rich, 8093, Switzerland}

\author[0000-0003-2871-127X]{Jorryt Matthee}
\affiliation{Department of Physics, ETH Z{\"u}rich, Wolfgang-Pauli-Strasse 27, Z{\"u}rich, 8093, Switzerland}
\affiliation{Institute of Science and Technology Austria (IST Austria), Am Campus 1, Klosterneuburg, Austria}

\author[0000-0001-9044-1747]{Daichi Kashino}
\affiliation{National Astronomical Observatory of Japan, 2-21-1 Osawa, Mitaka, Tokyo 181-8588, Japan}
\affiliation{Institute for Advanced Research, Nagoya University, Nagoya 464-8601, Japan}

\author[0000-0002-3120-7173]{Rongmon Bordoloi}
\affiliation{Department of Physics, North Carolina State University, Raleigh, 27695, North Carolina, USA}

\author[0000-0002-6423-3597]{Simon J.~Lilly}
\affiliation{Department of Physics, ETH Z{\"u}rich, Wolfgang-Pauli-Strasse 27, Z{\"u}rich, 8093, Switzerland}

\author[0000-0003-2895-6218]{Rohan P. Naidu}
\altaffiliation{NHFP Hubble Fellow}
\affiliation{MIT Kavli Institute for Astrophysics and Space Research, 77 Massachusetts Ave., Cambridge, MA 02139, USA}




\begin{abstract}
We report {\em JWST}/NIRCam measurements of quasar host galaxy emissions and supermassive black hole (SMBH) masses for six quasars at $5.9<z<7.1$ in the \textit{Emission-line galaxies and Intergalactic Gas in the Epoch of Reionization} (EIGER) project. We obtain deep NIRCam imaging in the F115W, F200W, and F356W bands, as well as F356W grism spectroscopy of the quasars. We use bright unsaturated stars to construct models of the point spread function (PSF) and estimate the errors of these PSFs. We then measure or constrain the fluxes and morphology of the quasar host galaxies by fitting the quasar images as a point source plus an exponential disk. We successfully detect the host galaxy of three quasars, which have host-to-quasar flux ratios of $\sim1\%-5\%$. Spectral Energy Distribution (SED) fitting suggests that these quasar host galaxies have stellar masses of $M_*\gtrsim10^{10}M_\odot$. For quasars with host galaxy non-detections, we estimate the upper limits of their stellar masses. We use the grism spectra to measure the {\hb} line profile and the continuum luminosity, then estimate the SMBH masses for the quasars. Our results indicate that the positive relation between SMBH masses and host galaxy stellar masses already exists at redshift $z\gtrsim6$. The quasars in our sample show a high black hole to stellar mass ratio of $M_\text{BH}/M_*\sim0.15$,  \textcolor{black}{which is about $\sim2$ dex higher than local relations. We find that selection effects only contribute partially to the high $M_\text{BH}/M_*$ ratios of high-redshift quasars. This result hints at a possible redshift evolution of the $M_\text{BH}-M_*$ relation.}
\end{abstract}



\section{Introduction}

Supermassive black holes (SMBHs) are ubiquitously found in the centers of galaxies \citep[e.g.,][]{kormendy01,tremaine02,heckman14}.
Observations of local galaxies have found tight correlations between SMBHs and the properties of their host galaxies, such as stellar masses or velocity dispersions, known as the $M_\text{BH}-M_*$ and the $M_\text{BH}-\sigma$ relations \citep[e.g.,][]{kh13}. 
These relations suggest a strong co-evolution between SMBHs and their host galaxies, likely through feedback during the active galactic nuclei (AGN) phases \citep[e.g.,][]{merloni08,ciotti01,ciotti10,fiore17}. Specifically, AGN activities can produce strong outflows and expel the cold gas content in their host galaxies, quenching the star formation and also exhausting the gas supply to the SMBH \citep[e.g.,][]{fabian12,cicone14,king15}.
Another important feedback mechanism is that AGNs can inject energy into the galaxy haloes through radio jets, which prevents halo gas from cooling and thereby shutting down both star-formation and BH accretion \citep[e.g.,][]{fabian12,heckman14}.

Meanwhile, other mechanisms might also contribute to these observed relations.
For example, 
SMBH accretion and galaxy star formation can be triggered by the same process \citep[e.g., galaxy mergers;][]{dm05,hopkins08}, producing a correlated growth of the SMBH and its host galaxy \citep[e.g.,][]{croton06,sb19}.
It has also been proposed that the SMBH-host correlation is a statistical effect resulting from the central limit theorem \citep[e.g.,][]{peng07,jahnke11}.
Given the debate of the possible scenarios, the exact origins of the $M_\text{BH}-M_*$ and the $M_\text{BH}-\sigma$ are still unclear. One critical unresolved question is whether these correlations are already established in the early universe at very high redshifts or if they gradually take shape throughout cosmic time. 

In the past two decades, about 300 quasars at $z>6$ have been discovered,
which indicates that SMBHs with $M_\text{BH}\gtrsim10^9M_\odot$
 already exist when the universe is less than 1 Gyr old
 \citep[e.g.,][]{mortlock11, jiang16, matsuoka18, wang19, yang20, wang21, banados23}.
 This quasar sample enables studies of SMBH-host co-evolution in the early universe. 
 By measuring the properties of these quasar host galaxies, we can characterize the $M_\text{BH}-M_*$ and the $M_\text{BH}-\sigma$ relations at $z\gtrsim6$.
So far, most of our knowledge about quasar host galaxies at $z\gtrsim6$ comes from sub-millimeter (sub-mm) wavelengths. Recent observations with the Atacama Large Millimeter Array (ALMA) have shown that quasars at $z\gtrsim6$ are hosted by massive starburst galaxies \citep[e.g.,][]{venemans16,decarli18,izumi19,yue21}. 
These quasar host galaxies have sizes of $\sim2-4$ kpc in the sub-mm and show a variety of morphologies and kinematics, 
ranging from rotation-dominated disks to dispersion-dominated irregular mergers \citep[e.g.,][]{venemans20, neeleman21}. 

Nevertheless, one important missing piece in our knowledge is the stellar component of high-redshift quasar host galaxies. 
The $M_\text{BH}-M_*$ and $M_\text{BH}-\sigma$ relations of local galaxies describe the connection between the black holes and the stellar components of their host galaxies; however, ALMA observations trace the emission from cold dust and gas, making it hard to make direct comparisons between the above mentioned ALMA observations and the local relations.
As luminous quasars are usually several magnitudes brighter than their host galaxies in rest-frame optical, probing the stellar emission of quasar host galaxies at $z\gtrsim6$ is extremely challenging. One viable way to measure the emission from quasar host galaxies is image decomposition, utilizing the fact that quasars appear to be point sources and their host galaxies are extended. This approach requires sharp point spread functions (PSFs) to disentangle the flux from the quasar and its host galaxy. Although image decomposition has been successful for quasars at \textcolor{black}{$z\lesssim2$} \citep[e.g.,][]{mechtley16,ding20,chen23},
detecting quasar host galaxies at $z\gtrsim6$ is much more challenging, given that the surface brightness of extended objects scales with redshift as $(1+z)^{-4}$.
Even with the sharp PSF of the {\em Hubble Space Telescope (HST)}, 
previous studies have only reached non-detections of quasar host galaxies at $z\gtrsim6$ \citep[e.g.,][]{marshall20}.

This situation was completely changed, however, by the recent launch of the {\em James Webb Space Telescope (JWST)}. 
With its $6.4-$meter aperture and infrared coverage, {\em JWST} provides even sharper PSF than the {\em HST} 
and is sensitive to the rest-frame optical emission of high-redshift quasar host galaxies.
An early study by \citet{ding23} reported the first detection of two quasar host galaxies at $z>6$ using NIRCam imaging, indicating that the two quasar hosts are among the most massive galaxies at their redshifts $(M_*\gtrsim10^{10}M_\odot)$.
\citet{marshall23} characterized the H$\beta$ and {\oiii} emission lines of two quasar host galaxies at $z\sim6.8$ using NIRSpec IFU,
showing complicated structures and kinematics of these galaxies.
These exciting results motivate us to increase the sample of high-redshift quasars with host galaxy measurements in rest-frame optical,
which will set a critical step towards fully understanding the co-evolution between SMBHs and galaxies in the early universe.

In this paper, we report the measurement of the host galaxy emission and SMBH properties for six luminous quasars at $z\gtrsim6$, using deep NIRCam imaging and spectroscopy as part of the \textit{Emission-line galaxies and Intergalactic Gas in the Epoch of Reionization} (EIGER) project. 
We use the images to measure the fluxes and morphologies of the quasar host galaxies,
and use the grism spectra to measure the black hole masses for the quasars from the \hb\ emission line.
Based on these measurements,
we discuss the implication of the quasar host galaxies on the $M_\text{BH}-M_*$ relation in the reionization era.

This paper is organized as follows. Section \ref{sec:data} describes the observations and data reduction. Section \ref{sec:image} describes the PSF modeling and the image fitting method we use to detect the rest-frame optical emission of the quasar host galaxies. We present the grism spectra and the SMBH mass measurements of these quasars in Section \ref{sec:bh} and discuss the co-evolution between high-redshift SMBHs and their host galaxies in Section \ref{sec:coevolve}. We discuss our results in Section \ref{sec:discuss}, and summarize this paper in Section \ref{sec:sum}. Throughout this paper, we assume a flat $\Lambda$CDM cosmology with $\Omega_M=0.3$ and $H_0=70\text{ km s}^{-1}\text{kpc}^{-1}$. All magnitudes are AB magnitudes unless further specified.

\section{Observations and Data Reduction} \label{sec:data}


The EIGER project (Proposal ID: 1243, PI: Lilly)
is a Guaranteed Time Observation (GTO) program targeting six quasars at redshift $5.9<z<7.1$,
delivering deep NIRCam imaging and wide-field slitless spectroscopy (WFSS)
of these quasar fields.
The information about these quasars is summarized in Table \ref{tbl:sample}.
This Section briefly describes the observations, and  
we refer the readers to \citet{eiger1}, \citet{eiger2}, \citet{eiger3}, and \citet{eiger4}
for more information about the EIGER project.

\begin{deluxetable*}{c|cccccc}
\label{tbl:sample}
\tablecaption{The quasar sample of the EIGER project}
\tablewidth{0pt}
\tablehead{\colhead{Quasar} & \colhead{RA} & \colhead{Dec} &  \colhead{Redshift} & \colhead{$M_{1450}$\tablenotemark{1}} &  \colhead{$\log M_\text{BH}^\text{\mgii}$\tablenotemark{2}} & \colhead{Reference\tablenotemark{3}} \\
\colhead{} & \colhead{(hh:mm:ss.ss)} & \colhead{(dd:mm:ss.s)} & \colhead{} & \colhead{(mag)} & \colhead{$(M_\odot)$} & \colhead{}}
\startdata
\hline
J0100+2802 & 01:00:13.02 & +28:02:25.80 & 6.327 & -29.02 & $10.1^{+0.2}_{-0.1}$ & \citet{xqr30}, \citet{Mazzucchelli23}\\
J0148+0600 & 01:48:37.64 & +06:00:20.0 & 5.977 & -27.08 & $9.58^{+0.08}_{-0.06}$ & \citet{xqr30}, \citet{Mazzucchelli23} \\
J1030+0524 & 10:30:27.11 & +05:24:55.06 & 6.304 & -26.99 & $9.27\pm0.09$ & \citet{xqr30}, \citet{Mazzucchelli23} \\
J159--02  &  10:36:54.19 & -02:32:37.94 & 6.381 & -26.47 & $9.49^{+0.049}_{-0.045}$ & \citet{banados16}, \citet{paolo22}\\
J1120+0641 & 11:20:01.48 & +06:41:24.3 & 7.085 & -26.44 & $9.13\pm0.01$ & \textcolor{black}{\citet{mortlock11}}, \citet{yang21}\\
J1148+5251 & 11:48:16.64 & +52:51:50.3 & 6.422 & -27.62 & $9.94\pm0.02$ & \citet{banados16}, \citet{shen19} \\\hline
\enddata
\tablenotetext{1}{The absolute magnitude at rest-frame $1450${\AA}.}
\tablenotetext{2}{The SMBH masses calculated using the {\mgii} broad emission line. The errors in the table only contain random errors; the systematic errors of the black hole mass is about 0.4 dex.}
\tablenotetext{3}{References for the quasar properties.}
\end{deluxetable*}

We obtain NIRCam F115W, F200W, F356W imaging, and F356W grism spectroscopy of the quasars. 
The observations of each quasar contain four individual visits,
forming a mosaic that covers a field of view (FoV) of $3'\times6'$ around the quasar.
See \citet{eiger1} for more information about the mosaic configuration.
The central $40''\times40''$ around each quasar is covered by every visits.
We adopt the INTRAMODULEX primary dither pattern and the 4-Point subpixel dither pattern
to improve the PSF sampling, cover detector gaps, and remove bad pixels.
By the time of writing this paper, the observation of four quasars are completed, 
while we only have two of the four visits yet for J159--02 and three visits for J1030+0524.
Our work presented here is based on these observations described above.
The exposure time per visit is 4,381s for the F115W imaging, 5,959s for the F200W imaging, 1,578s for the F356W imaging, and 8,760s for the grism spectroscopy. 

\subsection{NIRCam Imaging} \label{sec:data:image}


The NIRCam images were reduced using the {\texttt{jwst}} pipeline version 1.8.4.
We first run {\texttt{Detector1Pipeline}} to generate the rate files (*rate.fits),
then run {\texttt{Image2Pipeline}} to obtain calibrated images (*cal.fits).
For astrometry, we first align the calibrated images to each other using {\texttt{tweakwcs}}, 
then combine all the images and calibrate the absolute astrometry to the {\em Gaia} DR2 catalog \citep{gaiadr2}. 
We correct the $1/f$ noise, mask snowballs, subtract the wisp patterns, and remove cosmic rays for the images using custom codes
\citep[see ][for more detailed description]{eiger1}. 
We then run {\texttt{Image3Pipeline}} to stack images with the same visit and the same module.
We use a pixel size of $0\farcs03$ for the F356W images and $0\farcs015$ for the F115W and the F200W images. 

There are some noticeable differences between the image reduction of this work and that of previous EIGER papers.
Previous EIGER papers presented JWST imaging by combining all exposures in each field to form a final co-added image per filter. 
In this work, we only combine images with the same filter, visit, and module.
This approach reduces the systematic uncertainties in PSF modeling and image fitting (Section \ref{sec:image}) introduced by imperfect astrometric alignment between the visits.
Working on individual visits separately also allows us to estimate the systematic errors of the host galaxy measurements by comparing the results of different visits.
In addition, previous EIGER papers used a pixel scale of $0\farcs03$ for the stacked F115W and F200W images,
while in this work, we use a pixel scale of $0\farcs015$ to improve the sampling of the PSFs.

\subsection{NIRCam WFSS}



We obtain the NIRCam grism spectroscopy of the quasar fields using the grism ``R" in the F356W filter. This configuration gives an observed wavelength range of $3.1\mu\text{m}<\lambda<4.0\mu\text{m}$, which covers the {\hb} emission line at $5.8<z<7.2$. The data was reduced using {\texttt{jwst}} pipeline version 1.8.5. For each quasar, we first run the {\texttt{Detector1pipeline}} to obtain the *rate.fits files, then assign the world coordination system (WCS) information to the exposures using the {\texttt{AssignWcsStep}} and apply flat fielding using the {\texttt{FlatFieldStep}}. We remove $1/f$ noise and sky background variations by subtracting the median value in each column. We then trace and extract the 2D spectra of the quasar from individual exposures using custom scripts utilizing the {\texttt{grismconf}} module \citep[see][for more details]{kashino23}. Finally, we extract the spectra from all exposures using optimal extraction \citep{horne86}, and combine the extracted 1D spectra using the coadd1d pipeline in the PypeIt package \citep{pypeit}. 

\section{Measuring the Emission of Quasar Host Galaxies} \label{sec:image}


We use the NIRCam images to measure the flux and morphology of the quasar host galaxies. 
To do this, we fit the images of a quasar as a point source for the AGN plus an exponential disk for the host galaxy,
and use the best-fit parameters of the exponential disk to infer properties of the quasar host galaxy.

\subsection{PSF Models and Errors} \label{sec:image:psf}

\begin{figure*}
    \centering
    \includegraphics[width=\linewidth, trim={0.3cm 0 0.3cm 0}, clip]{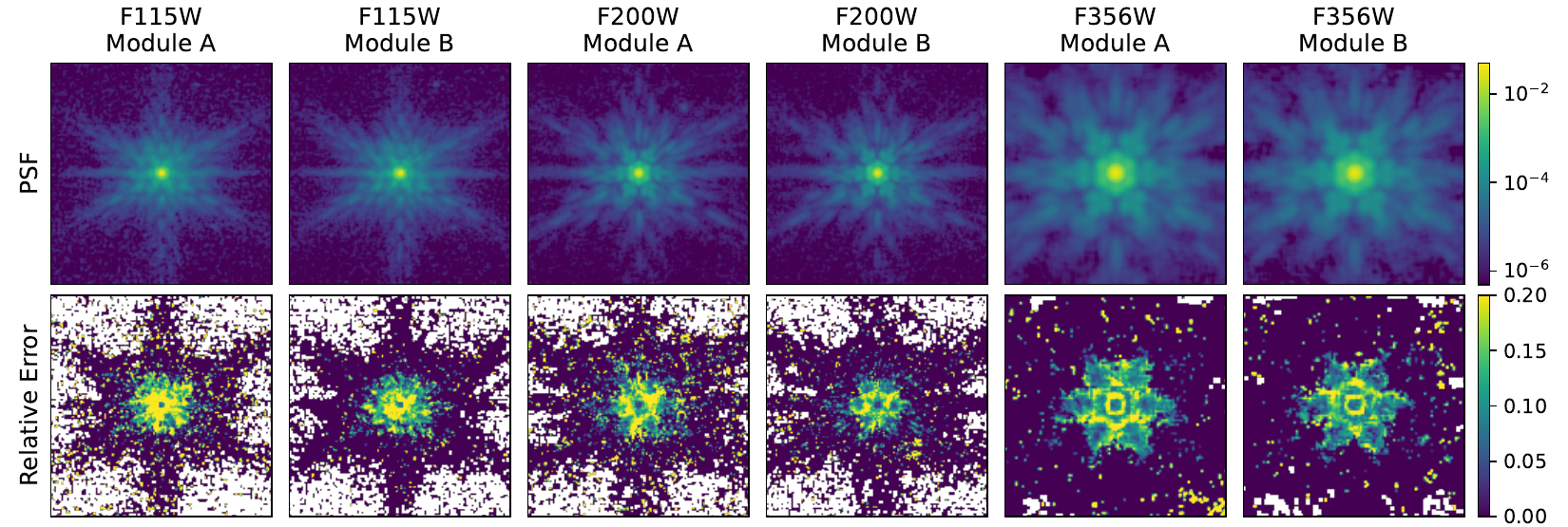}
    \caption{The PSFs and the relative error maps, estimated from isolated bright stars in the images. These cutouts have sizes of $3''\times3''$, and the integrated fluxes of the PSFs are normalized.
    The relative error is larger for brighter pixels, and the central bright pixels have relative errors up to $\sim30\%$.}
    \label{fig:psfs}
\end{figure*}

We construct PSF models using bright and unsaturated stars in the NIRCam images. This method has been found to provide accurate PSF models for quasar host galaxy detection and outperforms {\texttt{webbpsf}} \citep[e.g.,][]{ding22, zhuang23}. In this work, we use {\texttt{photutils}}\footnote{https://photutils.readthedocs.io/en/stable/index.html} to build effective PSFs, which uses the algorithm described in \citet{anderson00}. The detailed approach is as follows.

For each image, we first perform a source detection using the {\texttt{DAOphot}} algorithm \citep{daophot}, then select  
objects with magnitudes $18<m<21$ in each filter
and full-width half maxima (FWHMs) consistent with point sources\footnote{The specific FWHM limits are $(0\farcs0569, 0\farcs0695)$ for F115W, $(0\farcs0729, 0\farcs0789)$ for F200W, and $(0\farcs128,0\farcs152)$ for the F356W filter. These values correspond to the $3\sigma$ limits as measured by \citet{zhuang23}} as PSF stars. 
The magnitude cut is determined to match the fluxes of the quasars and avoid saturation, 
as the PSF shapes of infrared detectors exhibit flux dependence \citep[e.g., the brighter-fatter effect;][]{plazas18}.
We visually inspect all the PSF stars and reject those with close companions or bad pixels. 
We then use the {\texttt{EPSFBuilder}} class in the {\texttt{photutils}} package to build the empirical PSFs. 
Since the PSF of NIRCam depends  on the position on the focal plane \citep[e.g.,][]{zhuang23}, 
we construct the PSF models for module A and module B separately.
We also notice that the number of suitable PSF stars in a single image is very limited (usually fewer than three),
and some images do not have suitable PSF stars.
We thus include PSF stars from all quasar fields and visits when fitting the PSF models.
The typical number of PSF stars available for one filter and one module is $\sim10-20$. 

\textcolor{black}{It is worth noting that the PSF stars for the three bands are selected independently. Specifically, a PSF star in one band might be too bright or too faint to be selected as a PSF star in the other bands. As a result, the number of PSF stars available for the three bands are different.}

The empirical PSF models described above represent the average PSF of all the images. Limited by the number of PSF stars available,
we are not able to model the spatial and temporal variations of the PSFs.
Instead, we calculate the error maps of the PSF models, which estimate the possible differences between the PSF model and the real PSF of the quasar image.

Specifically, we first compute the differences between the flux-normalized images of the PSF stars (denoted by $P_i(x,y)$) and the PSF model (denoted by $\bar{P}(x,y)$),
then calculate the PSF error at pixel $(x,y)$ as
\begin{equation} \label{eq:psferr}
    \epsilon_P(x,y)^2=\frac{1}{N-1}\sum_i^N {[P_i(x,y)-\bar{P}(x,y)]^2-\sigma_i(x,y)^2}
\end{equation}
where $\sigma_i(x,y)$ is the random noise (i.e., the ERR extension of the images) of the $i-$th PSF star.
In practice, we perform sigma clipping with $\sigma_\text{limit}=3$ for each pixel in order to reduce the impact of outliers.
The error map gives the standard deviation of the pixels in the PSF models and is added to the noise map in the image fitting (Section \ref{sec:image:fitting}).

The final product of the PSF modeling step contains six PSFs (three filters times two modules).
The PSFs have sizes of $3''\times3''$ and have the same pixel size as the images. Figure \ref{fig:psfs} shows the PSF models and their relative error maps. 
There is a general trend that brighter pixels have larger relative errors;
the central pixels of the PSFs have relative errors up to $\sim30\%$. \textcolor{black}{The mean relative errors of the PSFs within radii of $3\times$FWHM are 0.204, 0.158, and 0.136 for the F115W, the F200W, and the F356W bands. This result is consistent with the finding in \citet{zhuang23}, who suggested that the errors of NIRCam PSF decrease toward long wavelengths.}

All PSF models, the PSF star lists, and the code to construct the effective PSF with error maps will be made available online upon publication.

\subsection{Image Fitting} \label{sec:image:fitting}

We use a point source component to describe the quasar and an exponential profile (i.e., a S\'ersic profile with index $n=1$) to describe the quasar host galaxy. 
We also add a constant background component to model imperfect background subtraction and add additional S\'ersic profiles   when there are other bright galaxies close to the quasar.
We use {\texttt{psfMC}} \citep{mechtley16} to perform image fitting, which is a Python-based package explicitly designed for quasar host galaxy detections utilizing the Markov chain Monte Carlo (MCMC) method.
We assign flat priors to all the free parameters, namely the magnitudes and positions of the point sources and S\'ersic profiles, the half-light major and minor radii and the position angle of the S\'ersic profiles,  the S\'ersic index of the S\'ersic profiles (except for the quasar host galaxy which is forced to have $n=1$), and the background level. 

\textcolor{black}{We fix the S\'ersic indices of the quasar host galaxies, which can only be poorly constrained due to the errors of the PSF models. The choice of $n=1$ (i.e., exponential disks) is based on the following motivations: (1) previous detections of high-redshift quasar host galaxies are consistent with exponential profiles \citep{ding23}; (2) high-resolution ALMA observations found that the far-infrared emissions of quasar host galaxies have S\'ersic index $n\approx1$ \citep[e.g.,][]{yue21}. We will discuss more about this point at the end of this Section.}

It is worth noting that \texttt{psfMC} incorporates PSF error maps when evaluating the likelihood of an image model. Specifically, the error of a pixel is calculated by combining the PSF error and the random noise of the image, i.e.,
\begin{equation} \label{eq:adderr}
\epsilon_\text{all}^2(x,y) = F_P^2\epsilon_P^2(x,y) + \sigma^2(x,y)
\end{equation}
where $\epsilon_\text{all}(x,y)$ stands for the combined error at pixel $(x,y)$, $F_P$ is the flux of the PSF, 
$\epsilon_P(x,y)$ is the \textcolor{black}{PSF error at pixel $(x,y)$} (described in Section \ref{sec:image:psf}),
and $\sigma(x,y)$ is the random noise of the pixel. 
By including the PSF error map in the total error, we assign lower weights to pixels with large PSF uncertainties and reduce their impact on the fitting result. 
This combined error map also helps us to distinguish quasar host galaxy emissions from PSF inaccuracies in the PSF-subtracted images.


For each quasar, we first fit its F356W images to determine whether the host galaxy is detected and to measure the morphology of the host galaxy.
This choice is made \textcolor{black}{based on two considerations. First, long-wavelength NIRCam filters exhibit smaller spatial variation, \textcolor{black}{i.e., NIRCam PSFs are more stable at longer wavelengths, as we discuss in Section \ref{sec:image:psf}.} Second, the flux ratio between the host galaxy and the quasar $(F_G/F_Q)$ increases towards longer wavelength \citep{marshall21}. The reason is twofold: (1) quasars have blue continua in the rest-frame optical; (2) the F356W filter probes wavelengths longer than the 4000{\AA}-break for $z\sim6-7$ galaxies. As we will later show in this Section, the quasars with host galaxy detections in our sample have $(F_G/F_Q)_\text{F356W}\sim1-3\times(F_G/F_Q)_\text{F200W}$.}

\textcolor{black}{
We fit the F356W images from individual visits separately, and
determine the best-fit parameters and their errors by computing the median and standard deviation of the MCMC samples from all the visits.
We also fit the images using a single PSF plus a sky background to evaluate the improvement in reduced $\chi^2$ by including a host galaxy component in the model. Specifically, we use the relative improvement of the reduced $\chi^2$,
\begin{equation} \label{eq:rchisq}
    \frac{\Delta \chi^2_\nu}{\chi^2_\nu}=\frac{\chi^2_\nu(1p)-\chi^2_\nu(1p1e)}{\chi^2_\nu(1p1e)}
\end{equation}
where $\chi^2_\nu(1p)$ is the reduced $\chi^2$ for the single-PSF model, and $\chi^2_\nu(1p1e)$ represents the reduced $\chi^2$ where the exponential disk component is added to the model.}
\textcolor{black}{A quasar host galaxy is considered to be detected in the F356W band if the fitting result meets the following criteria:
\begin{enumerate}
    \item The error of the quasar host galaxy's magnitude is smaller than 0.3 (i.e., a detection with $>3\sigma$ significance);
    \item The half-light radius satisfies $R_{e, \text{maj}}<0\farcs9$ and $R_{e, \text{min}}>0\farcs1$;
    \item The axis ratio satisfies $0.3<R_{e, \text{min}}/R_{e, \text{maj}}<1$;
    \item All visits have $\Delta \chi^2_\nu/\chi^2_\nu>0.1$.
\end{enumerate}}
\textcolor{black}{Here, criterion (1) ensures that the host galaxy is consistently detected in all visits; criterion (2) ensures that the best-fit host galaxy has a reasonable size and is unlikely to be confused with the PSF component or the background component; criterion (3) further excludes some false detections caused by PSF inaccuracies; criterion (4) ensures that the quasar's image is significantly different from a single PSF. We will discuss why we adopt these criteria in Appendix \ref{sec:image:fitting:limit} with more details.}

\textcolor{black}{
For quasars with successful host galaxy detections in the F356W band, 
we fit the F200W and the F115W images by fixing the position and the morphology of the host galaxy to the best-fit values from the F356W images. Again, we fit the images from different visits separately,
and estimate the best-fit parameters and their errors using the median and the standard deviation of the MCMC samples from all the visits.
We report a host galaxy detection in the F200W or the F115W band if the quasar has $\Delta \chi^2_\nu/\chi^2_\nu>0.07$ and has a host galaxy magnitude error smaller than 0.3 mag.
For quasars with non-detections in the F356W band, we fit their F200W and F115W images as a point source plus a constant background.}

\textcolor{black}{We notice that the morphology and positions of quasar host galaxies might be wavelength-dependent. Fixing the morphology and positions of quasar host galaxies when fitting the F200W and the F115W band images might introduce additional systematic errors. However, leaving these parameters free will return non-detections of quasar host galaxies in most cases, due to the fainter quasar host galaxies and the stronger PSF errors at short wavelengths. Other observations of high-redshift quasars find that the morphology of quasar host galaxies is consistent in the SW and LW bands \citep{ding23}. As such, we determine that fixing the host galaxy morphology and position to the best-fit F356W values is the best approach based on our data.}



\begin{figure*}
    \centering
    \includegraphics[width=0.44\linewidth,trim={0 0.8cm 0 0}]{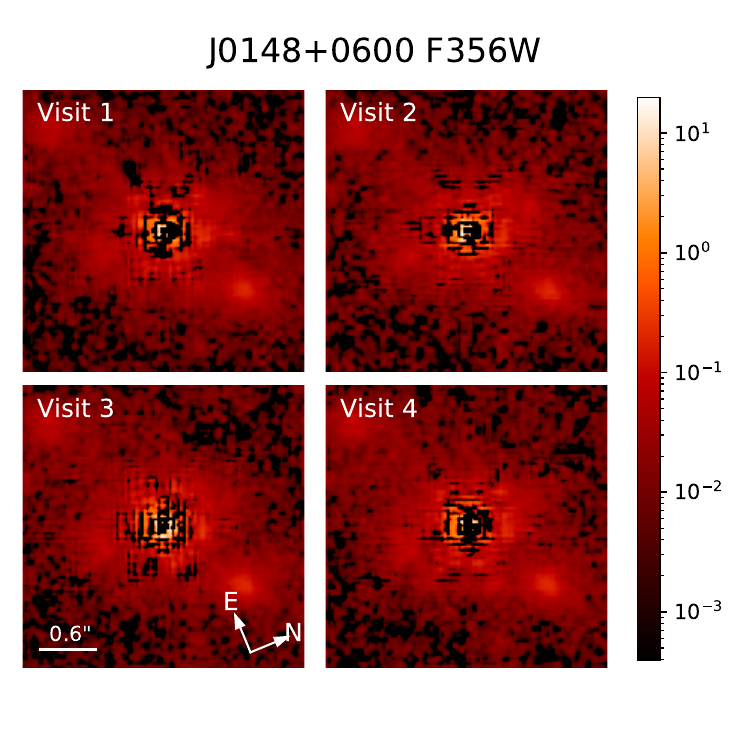}
    \includegraphics[width=0.55\linewidth]{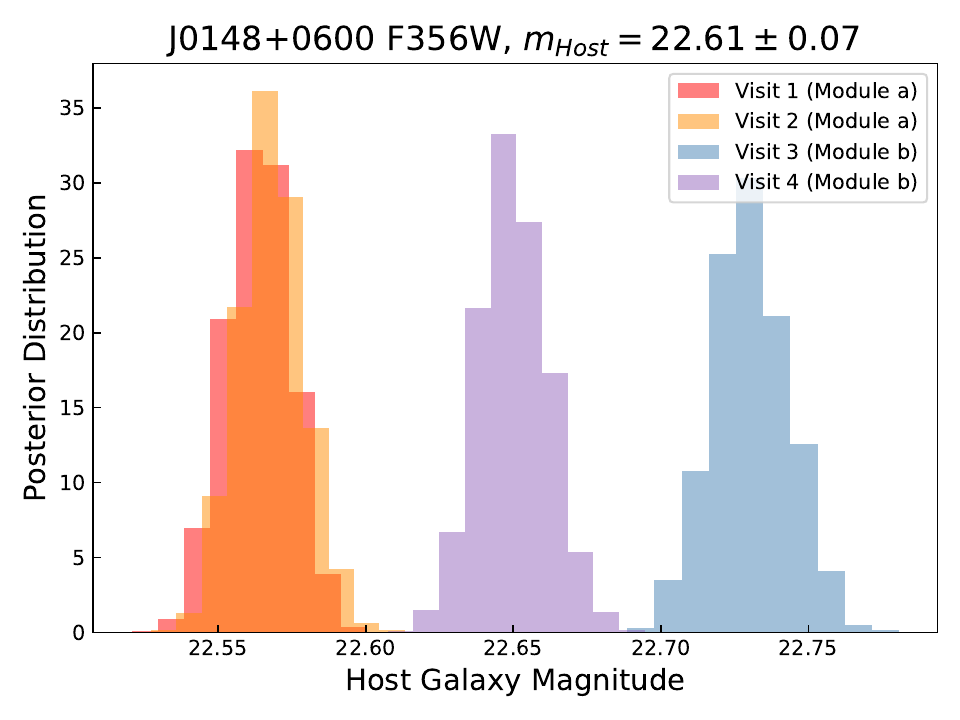}
    \caption{Fitting the F356W images of J0148+0600 from four visits. This Figure illustrates how we validate the detections of host galaxies. {\em Left}: the PSF-subtracted images, which show consistent shapes and brightness across the visits. {\em Right}: the MCMC posterior distribution of the host galaxy magnitudes. The small inter-visit differences reflect the systematic uncertainties of the host galaxy flux measurements. \textcolor{black}{We take the median of all the MCMC samples from all four visits as the best-fit host galaxy magnitude, and
    estimate its uncertainty by computing the standard deviation of all the MCMC samples.}}
    \label{fig:J0148visits}
\end{figure*}

We demonstrate how we validate the host galaxy detections and estimate the uncertainties of host galaxy properties in Figure \ref{fig:J0148visits}.
The left panel of Figure \ref{fig:J0148visits} shows the PSF-subtracted F356W image of J0148+0600 from individual visits.
We emphasize that the quasar is located in module A (B) in visits 1 and 2 (3 and 4), 
and we model the PSF of module A and module B independently. As such,
the similar patterns in the PSF-subtracted images from individual visits strongly indicate that the host galaxy detection is reliable.
The right panel of Figure \ref{fig:J0148visits} shows the host galaxy fluxes measured from the four visits. 
The histograms represent the posterior distributions from MCMC,
and the inter-visit differences reflect the systematic errors of image fitting.
We estimate the uncertainties of the host galaxy magnitude by computing the standard deviation of MCMC samples from all the visits, which
 takes into account both random errors and systematic errors.
 In this case, the standard deviation of the host galaxy magnitude is $\Delta m=0.07$, and the host galaxy is successfully detected.

\begin{deluxetable*}{c|ccccccccc}
\label{tbl:imagefitting}
\centering
\tablecaption{The Results of Image Fitting}
\tablewidth{0pt}
\tablehead{\colhead{Quasar} & \colhead{$m^\text{QSO}_\text{F115W}$} & \colhead{$m^\text{QSO}_\text{F200W}$} &  \colhead{$m^\text{QSO}_\text{F356W}$} & \colhead{$m^\text{host}_\text{F115W}$\tablenotemark{1}} & \colhead{$m^\text{host}_\text{F200W}$} & \colhead{$m^\text{host}_\text{F356W}$} & \colhead{$R_{e, \text{circ}}$} & \colhead{$e$\tablenotemark{2}} & \colhead{PA\tablenotemark{3}}\\
\colhead{} & \colhead{} & \colhead{} & \colhead{} & \colhead{} & \colhead{} & \colhead{} & \colhead{$('')$} & \colhead{} & \colhead{(deg)}
}
\startdata
\hline
J0148+0600 & $19.522\pm0.003$ & $18.912\pm0.001$ & $19.109\pm0.003$ & $23.48\pm0.24$ & $23.51\pm0.15$ & $22.61\pm0.07$ & $0.39\pm0.02$ & $0.37\pm0.04$ & $106\pm1$\\
J159--02  & $20.146\pm0.003$ & $19.680\pm0.002$ & $19.543\pm0.003$ & $(24.83\pm0.06)$ & $24.82\pm0.23$ & $23.98\pm0.16$ & $0.48\pm0.03$ & $0.27\pm0.05$ & $132\pm5$\\
J1120+0641 & $20.366\pm0.003$ & $19.886\pm0.002$ & $19.632\pm0.003$ & $(24.78\pm0.12)$ & $24.43\pm0.10$ & $24.45\pm0.20$ & $0.32\pm0.02$ & $0.57\pm0.12$ & $86\pm5$\\\hline
J0100+2802 & $17.87\pm0.01$ & $17.275\pm0.003$ & $17.172\pm0.001$ & - & - & - & - & - & - \\
J1030+0524 & $19.969\pm0.003$ & $19.514\pm0.001$ & $19.415\pm0.003$ & - & - & - & - & - & - \\
J1148+5251 & $19.140\pm0.001$ & $18.782\pm0.001$ & $18.782\pm0.002$ & - & - & - & - & - & - \\\hline
\enddata
\tablenotetext{1}{The F115W magnitudes of the host galaxies \textcolor{black}{for J159--02 and J1120+0641} are tentative detections. See section \ref{sec:image:fitting} and Appendix \ref{sec:image:fitting:limit}  more details.}
\tablenotetext{2}{The ellipticity is defined as $e=1-R_{e,\text{min}}/R_{e,\text{max}}$.}
\tablenotetext{3}{The position angle is defined as the angle between the major axis and the north vector, with counter-clockwise being positive.}
\tablecomments{The image fitting results of the quasars. The top three rows summarize the three quasars with host galaxy detections, and the bottom three rows give quasars with non-detections of the host galaxies. All errors are $1-\sigma$ errors.}
\end{deluxetable*}

Our image fitting procedure takes advantage of the multi-band, multi-visit observations of the EIGER project.
The uncertainties of quasar host galaxy measurements are dominated by systematic errors of the PSF model instead of \textcolor{black}{random photon  noise}.
Fitting the four visits individually allows us to validate the result and estimate the systematic uncertainties by comparing the output of all the visits.

\begin{figure*}
    \centering
    \includegraphics[width=1\linewidth,trim={6cm 2cm 4.5cm 0},clip]{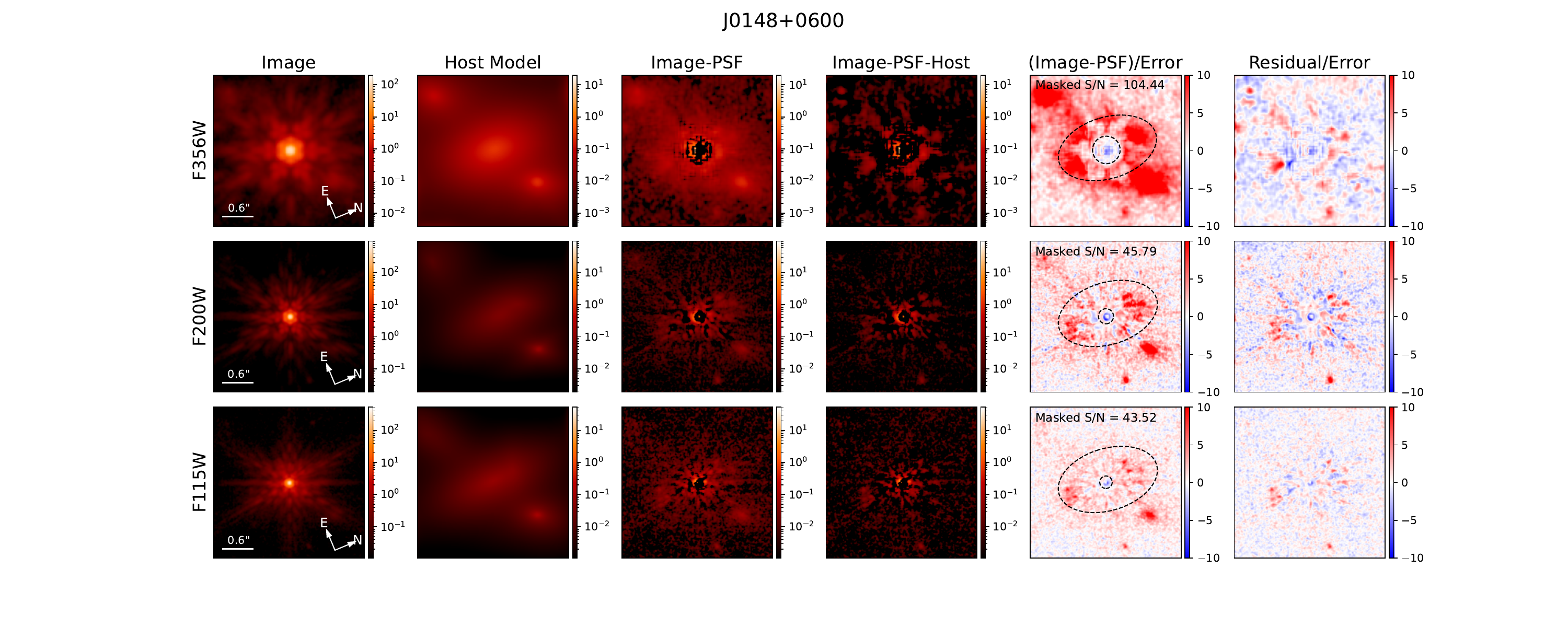}
    \caption{The image fitting results of J0148+0600. From left to right: the original NIRCam image, the host galaxy model, the PSF-subtracted image, the residual image, \textcolor{black}{the PSF-subtracted image divided by the composite error, and the residual image divided by the composite error}. \textcolor{black}{We obtain these images by combining images from all the visits.} We detect the host galaxy in all three bands. 
     {The dashed ellipses mark the apertures corresponding to twice of the half-light radii, and the dashed circles mark the central regions with radii of $2\times \text{FWHM}_\text{PSF}$. The ``masked S/N'' quoted here are the S/N of the host galaxy signal within the ellipse (excluding the central regions); see Appendix \ref{sec:image:fitting:limit} for more details.}
    J0148+0600 have two close companions, 
    which are modeled as S\'ersic profiles when fitting the images. Note that (1) the compisite error is a combination of random noises and PSF errors (see Equation \ref{eq:adderr}); (2) the ``Host Model" images include the two close companions of the quasar.}
    \label{fig:J0148images}
\end{figure*}
\begin{figure*}
    \centering
    \includegraphics[width=1\linewidth,trim={6cm 2cm 4.5cm 0},clip]{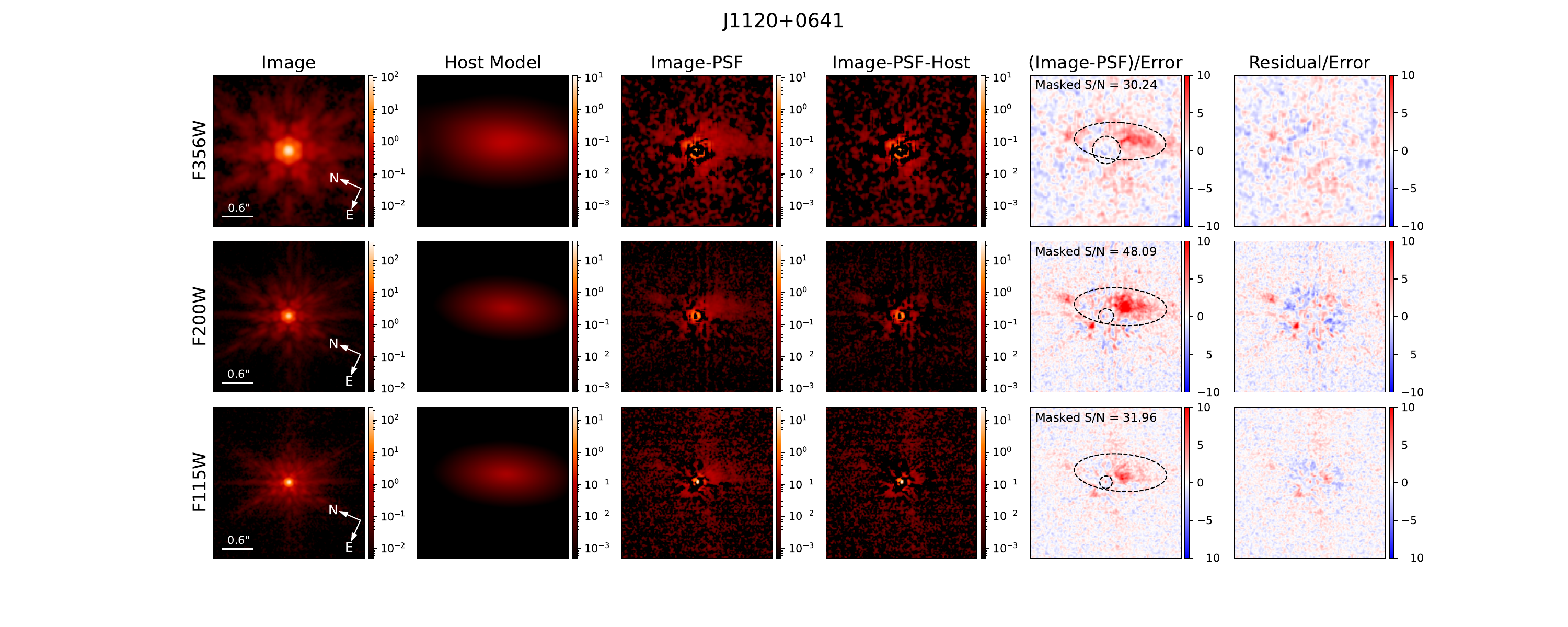}
    \caption{Same as Figure \ref{fig:J0148images}, but for J1120+0641. We detect the host galaxy in the F356W and the F200W band, and report a tentative detection in the F115W band. The PSF-subtracted images show consistent shapes in all three bands. We note that we exclude the F356W image from visit 4 due to a large number of bad pixels around the quasar. The host galaxy of J1120+0641 is about $0\farcs5$ away from the quasar, and the PSF-subtracted images exhibit irregular shapes. These features suggest that J1120+0641 might be hosted by an on-going merger. See Section \ref{sec:J1120} for more details.}
    \label{fig:J1120images}
\end{figure*}
\begin{figure*}
    \centering
    \includegraphics[width=1\linewidth,trim={6cm 2cm 4.5cm 0},clip]{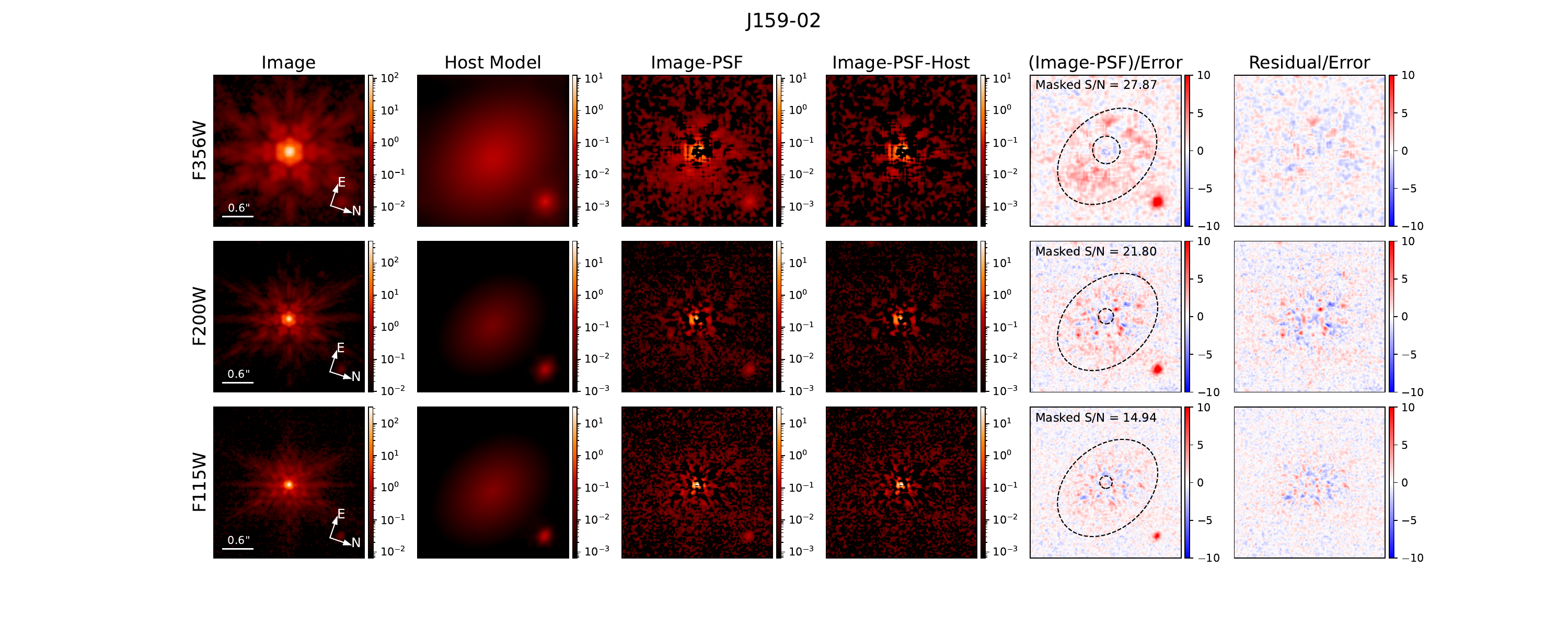}
    \caption{Same as Figure \ref{fig:J0148images}, but for J159-02. We detect the host galaxy in the F356W and the F200W band, and report a tentative detection in the F115W band. J159-02 has one close companion. Note that we only have two visits for J159-02 when writing this paper. See Section \ref{sec:J159} for more details.}
    \label{fig:J159images}
\end{figure*}
\begin{figure*}
    \centering
    \includegraphics[width=0.6\linewidth,trim={2cm 2cm 1cm 0},clip]{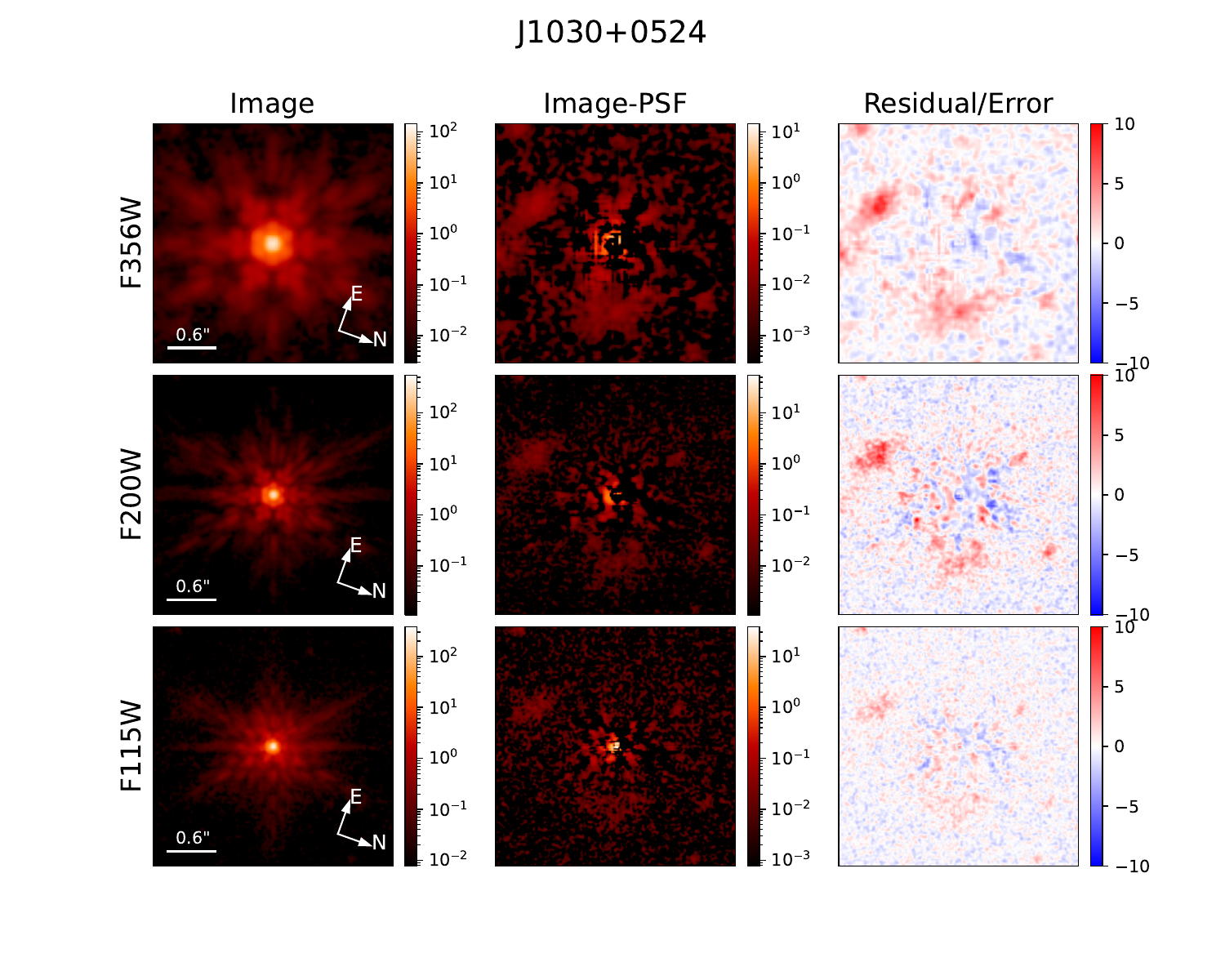}
    \caption{The image fitting results of J1030+0524. From left to right: the original image, the PSF-subtracted residual, the residual normalized by the image error.
    We do not detect the host galaxy according to the criteria in Section \ref{sec:data:image}. 
    The PSF-subtracted images exhibit extended emissions around the quasar, which might be tidal tails of a recent galaxy merger. Note that we only have three visits for J1030+0524 when writing this paper.}
    \label{fig:J1030images}
\end{figure*}
\begin{figure*}
    \centering
    \includegraphics[width=0.6\linewidth,trim={2cm 2cm 1cm 0},clip]{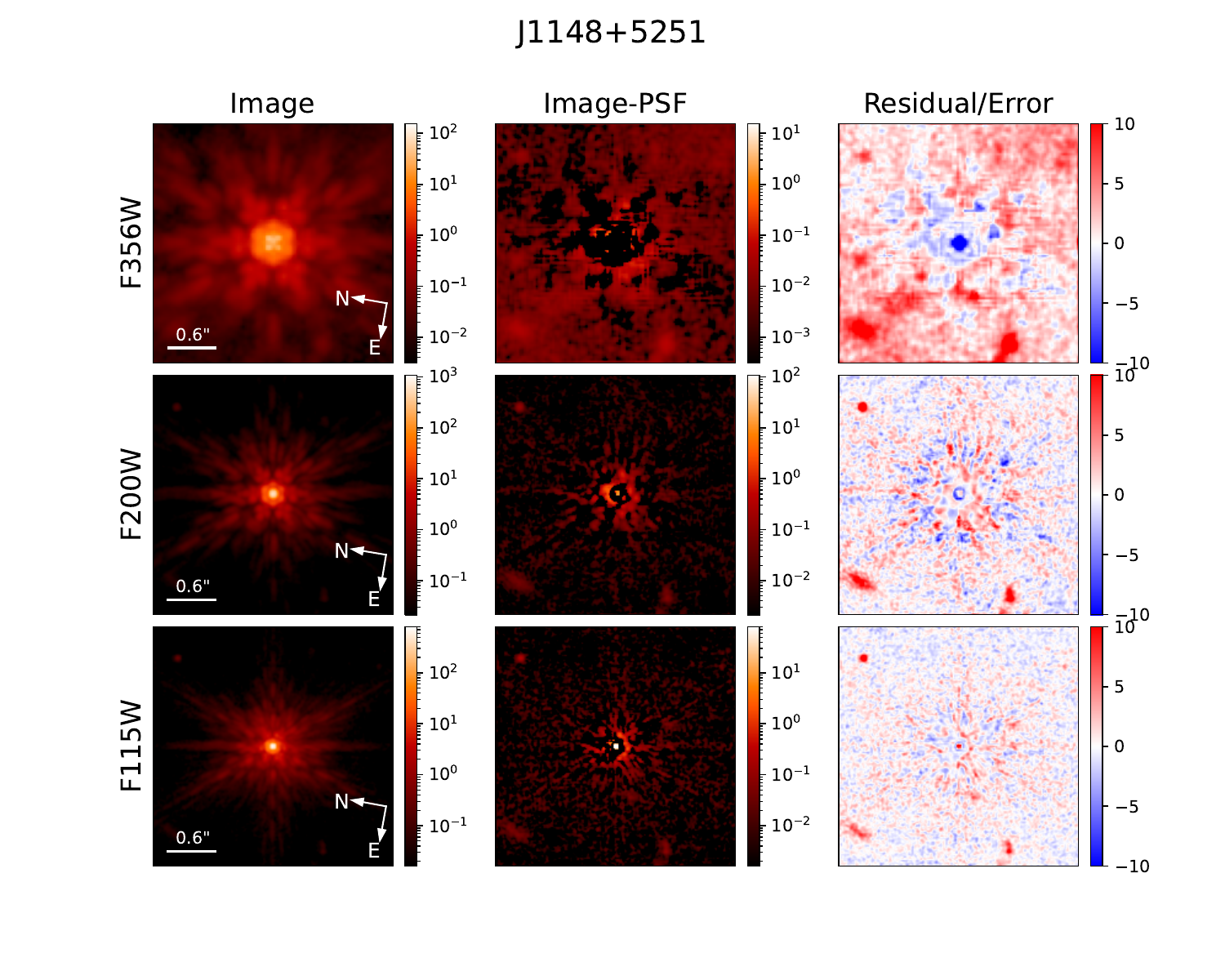}
    \caption{Same as Figure \ref{fig:J1030images}, but for J1148+5251. We do not detect the host galaxy according to the criteria in Section \ref{sec:data:image}. 
    The F356W image exhibits diffused emission extending from the lower left to the upper right corner, which is not seen in the other two bands. This feature might be {\oiii} emission around the quasar. We also see several close companions around the quasar that are detected in all three bands. \textcolor{black}{Note that the central pixels in the F356W band are saturated.}}
    \label{fig:J1148images}
\end{figure*}

Figure \ref{fig:J0148images} to \ref{fig:J1148images} present the results of the image fitting. 
We summarize the best-fit parameters in Table \ref{tbl:imagefitting}.
The PSF-subtracted images of these quasars exhibit a variety of features.
We successfully detect the host galaxy of \textcolor{black}{J0148+0600 in all the three bands, as well as J159--02 and J1120+0641 in the F356W and the F200W bands. J159--02 and J1120+0641 have $\Delta \chi^2_\nu/\chi^2_\nu\sim0.05$ in the F115W band, which do not satisfy the reduced $\chi^2$ criterion; nevertheless, we still report tentative detections for their host galaxies in the F115W band because the magnitude errors are smaller than 0.3 mag.}
The flux ratios between the host galaxies and the quasars range from $\sim1\%-5\%$.

\textcolor{black}{The ``(Image-PSF)/Error'' maps clearly show the impact of the PSF error. Pixels that have large PSF errors (mainly those at central regions and on PSF spikes) are assigned lower weights in image fitting, and their influence on the fitting results is suppressed. The regions where the PSF errors are small exhibit significant host galaxy emissions compared to the noise level, confirming that the host galaxy signal is real and is not a result of PSF inaccuracies. We leave a more detailed discussion on how we validate the host galaxy detections and why we adopt the detection criteria in Appendix \ref{sec:image:fitting:limit}.}

The host galaxies of J1030+0524 and J1148+5251 are not detected according to the detection criteria described above. However, the PSF-subtracted images of the two quasars clearly show extended emissions around the quasar, which might be tidal tails of a recent merger or the diffuse {\oiii} emission from galactic-scale outflows. 
The analysis of J0100 + 2802 has already been reported in \citet{eiger3}, and we refer the readers to \citet{eiger3} for more details. As a quick summary, J0100+2802 is saturated in all the images due to its extreme brightness, and we do not detect the host galaxy or any extended emission in its PSF-subtracted images.

We also try to fit the images without fixing the S\'ersic index of the host galaxy. We find that the S\'ersic indices of the host galaxies are poorly constrained. \textcolor{black}{Specifically, the posterior distribution of the indices spans a wide range at $0.3\lesssim n \lesssim 4$, with estimated errors $\sigma_n\gtrsim1$. We notice that the best-fit host galaxy magnitudes of the S\'ersic profiles are similar to the exponential disk models (with differences of $\lesssim0.3$ mag). This systematic uncertainty is much smaller than the errors of the stellar mass estimates ($\sim0.3$dex, see Section \ref{sec:image:sed}) and has no major influence on the main conclusions of this work.}

We provide more information about individual quasars in Section \ref{sec:image:individual}.


\subsection{SED fitting}  \label{sec:image:sed}

One of the main goals of this work is to characterize the position of luminous high-redshift quasars on the $M_\text{BH}-M_*$ plot.
To measure the stellar masses of the quasar host galaxies, we perform SED fitting for these galaxies using {\texttt{Prospector}} \citep{prospector},
with nebular emission treatments based on {\texttt{Cloudy}} \citep[see][for details]{byler17}.
\textcolor{black}{Since we only report tentative detections for the host galaxies of J159--02 and J1120+0641 in the F115W band,
we only take the F356W and the F200W magnitudes as the input of the SED fitting for these two quasars.}

We assume a delayed-$\tau$ model for the star formation history (SFH), 
i.e., $\text{SFR}(t)\propto te^{-t/\tau}$.
We use a Chabrier initial mass function \citep{chabrier03} and assume a dust attenuation following the \citet{calzetti03} law.
The free parameters and their priors of this SED model include: 
(1) the stellar mass $M_*$ with a log-uniform prior at $[10^8M_\odot, 10^{12}M_\odot]$;
(2) the stellar metallicity $\log (Z/Z_\odot)$ with a uniform prior at $[-2, 0.2]$;
(3) the starting time of the star formation $t_\text{age}$ with a uniform prior at $[0, t(z)]$, where $t(z)$ is the age of the universe at the quasar's redshift;
(4) the exponential decay timescale $\tau$ with a uniform prior at $[0.01\text{Myr}, 20\text{Myr}]$;
(5) the dust attenuation (quantified as the optical depth at 5500{\AA}, $\tau_{5500}$) with a uniform prior at $[0, 2]$; 
(6) the gas-phase metallicity $\log (Z_g/Z_\odot)$ with a uniform prior at $[-2, 0.5]$; 
and (7) the ionization parameter $\log U$ with a uniform prior at $[-3, 1]$. 
\textcolor{black}{With only two or three photometric points for the SED fitting, we are only able to constrain the stellar masses, which roughly give the normalization of the SED. All other parameters remain largely unconstrained.}

\begin{figure*}
    \centering
    \includegraphics[width=1\linewidth, trim={0.5cm 0 0 0}, clip]{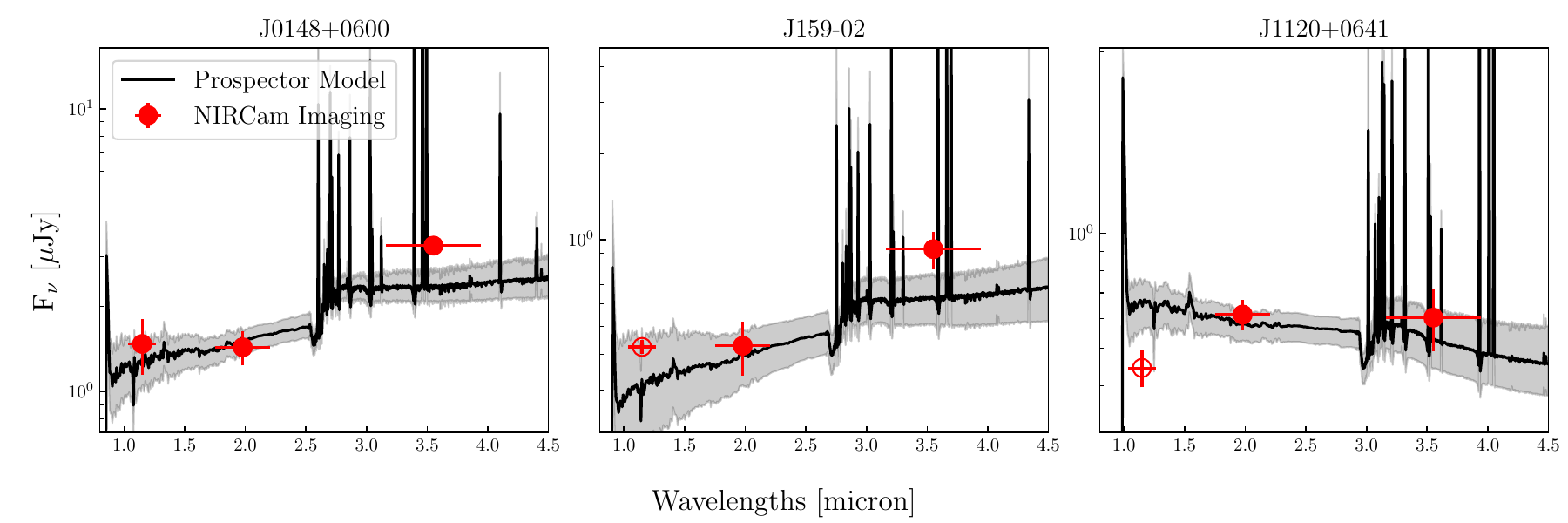}
    \caption{Fitting the SEDs of the quasar host galaxies. The solid red dots represent the F200W and the F356W magnitudes of the quasar host galaxies. The black line shows the median modeled spectra generated using {\texttt{prospector}}, and the shaded area gives the $1\sigma$ error. For comparison, \textcolor{black}{we also show the F115W magnitudes of the host galaxies of J159--02 and J1120+0641 in open circles, which are not used when fitting their SEDs}. The SED fitting enables constraints on the stellar masses of the quasar host galaxies.}
    \label{fig:sed}
\end{figure*}

We first perform the SED fitting for the detected quasar host galaxies.
Figure \ref{fig:sed} shows the best-fit SEDs, and Table \ref{tbl:properties} summarizes the physical parameters of the quasar host galaxies.
The quasar host galaxies detected in this work have stellar masses of $M_*\gtrsim10^{10}M_\odot$, which are among the most massive galaxies at $z\gtrsim6$.
\textcolor{black}{We also plot the tentative F115W detections of the host galaxies of J159--02 and J1120+0641 for reference, but do not include them in the SED fitting. 
The F115W fluxes of J159--02 is consistent with the best-fit SED model, while the F115W flux of J1120+0641 is $\sim2\sigma$ lower than the SED model.}

For the quasars with only non-detections of their host galaxies, we estimate the upper limits of their host galaxy stellar masses. 
Specifically, we set \textcolor{black}{conservative} lower limits of the host galaxy magnitudes as $m_\text{host} = m_\text{QSO}+3.5$ in the F356W band, then scale the best-fit host galaxy SED model of J0148+0600 to match this magnitude \textcolor{black}{and take the corresponding stellar masses as upper limits.} \textcolor{black}{This choice is made because J0148+0600 has $\Delta m=m_\text{host}-m_\text{QSO}=3.5$ in the F356W band and has a significant host galaxy detection.
We note that the three quasars with host galaxy non-detections have F356W magnitudes similar to or brighter than J0148+0600; if the host galaxies of these quasars have $\Delta m_\text{F356W}=3.5$, we expect that the host galaxies should be detected at significance levels similar to or higher than J0148+0600.}
The stellar mass upper limits of these quasar host galaxies are also listed in Table \ref{tbl:properties}, which have $M_*\sim10^{10.5}-10^{11.5}M_\odot$.



\begin{deluxetable*}{c|cccc|ccc}
\label{tbl:properties}
\tablecaption{The properties of the quasars and their host galaxies}
\tablewidth{0pt}
\tablehead{\colhead{Quasar} &  \colhead{$\text{FWHM}_\text{\hb}$} & \colhead{$\log L_{5100}$} & \colhead{$\log M_\text{BH}$\tablenotemark{1}} & \colhead{$\lambda_\text{Edd}$} & \colhead{$\log M_*$\tablenotemark{2}} & \colhead{$M^\text{host}_\text{UV}$\tablenotemark{3}}  & \colhead{$R_{e, \text{circ}}$}\\
\colhead{} & \colhead{$(\text{km s}^{-1})$} & \colhead{$(\text{erg s}^{-1})$}  & \colhead{$(M_\odot)$} & \colhead{} & \colhead{$(M_\odot)$} & \colhead{(mag)}  & \colhead{(kpc)}}
\startdata
\hline
J0148+0600 & $7828_{-480}^{+485}$ & $46.390_{-0.002}^{+0.001}$ & $9.892_{-0.055}^{+0.053}$ & 0.23 & \textcolor{black}{$10.64^{+0.22}_{-0.24}$} & $-22.81^{+0.48}_{-0.39}$ & $2.23\pm0.11$ \\
J159--02  & $3493^{+29}_{-20}$ & $46.199_{-0.001}^{+0.001}$ & $9.096_{-0.005}^{+0.007}$ & 0.93 & $10.14^{+0.34}_{-0.36}$ & $-21.68^{+0.54}_{-0.42}$ & $2.64\pm0.16$ \\
J1120+0641 & $3337^{+95}_{-111}$ & $46.246_{-0.028}^{+0.016}$ & $9.076_{-0.030}^{+0.029}$ & 1.08 & $9.81^{+0.23}_{-0.31}$ & $-22.63^{+0.17}_{-0.13}$ & $1.66\pm0.10$ \\\hline
J0100+2802 & $6045_{-20}^{+22}$ & $47.1776_{-0.0004}^{+0.0003}$ & $10.062_{-0.003}^{+0.003}$ & 0.96 & $<11.58$ & $>-24.90$ & - \\
J1030+0524 & $3669^{+16}_{-14}$ & $46.295_{-0.001}^{+0.001}$ & $9.187_{-0.003}^{+0.004}$ & 0.94 &  $<10.65$ & $>-22.59$ & - \\
J1148+5251 & $5370_{-68}^{+81}$ & $46.541_{-0.002}^{+0.002}$ & $9.640_{-0.010}^{+0.012}$ & 0.59 & $<10.93$ & $>-23.28$ & - \\\hline
\enddata
\tablenotetext{1}{The errors listed in the Table are random errors from MCMC. The systematic errors of $\log M_\text{BH}$ is $\sim0.3$dex is dominated by the scatter of the empirical relation (Equation \ref{eq:mbh}).}
\tablenotetext{2}{The errors are estimated from {\texttt{prospector}} posterior distributions. For quasars with non-detections, the upper limits correspond to $\Delta m_\text{F356W}=3.5$ (Section \ref{sec:image:sed}).}
\tablenotetext{3}{The absolute magnitude at rest-frame 1500{\AA} from SED fitting.}
\tablecomments{ All errors are $1-\sigma$ errors.}
\end{deluxetable*}

Since we only have \textcolor{black}{two or three} photometric points available, we are not able to adopt more complicated star formation histories (e.g., a non-parametric star formation history) in the SED fitting.
We also notice that the F356W fluxes of the quasar host galaxies contain both stellar continuum and the {\hb} and {\oiii} nebulae lines.
Since we do not have the nebulae line fluxes, we \textcolor{black}{leave all emission-line-related parameters free to account for this systematic uncertainty.}
Future observations with NIRSpec IFU will provide the fluxes of these nebulae lines, which will allow us to use more complicated star formation histories and improve the accuracy of the estimated stellar masses \citep[e.g.,][]{marshall23}.
\textcolor{black}{It is also possible to measure the extended line emissions from quasar host galaxies using the two-dimensional grism spectra of the quasar. However, such measurements require careful analysis of the two-dimensional grism data, which is beyond the scope of this paper.}

Finally, we perform a sanity check of the stellar mass estimates by comparing our results with mock galaxies in {\texttt{UniverseMachine}} Data Release 1 \citep{umdr1}. We select mock galaxies with redshifts $6<z<7$ and UV magnitudes $-23<M_\text{UV}<-21.5$, which roughly match the redshift and luminosity range of the quasar host galaxies. These mock galaxies have stellar masses of $10^{9.2}M_\odot-10^{10.7}M_\odot$ ($95\%$ confidence interval), with a median of $10^{9.96}M_\odot$.
These numbers are close to the stellar mass estimates for the quasar host galaxies in this work.

\section{Black Hole Mass Estimates for the Quasars} \label{sec:bh}

We calculate the black hole masses using the single-epoch virial estimator \citep[e.g.,][]{shen12}.
This method uses the FWHM of broad emission lines to estimate the velocity of the broad line region (BLR) clouds, 
and uses the continuum luminosity as a proxy of the distance from the BLR to the SMBH based on the luminosity-radius relation \citep[e.g.,][]{kaspi05}.
Specifically, the SMBH mass can be calculated using the following relation:
\begin{equation} \label{eq:mbh}
    \frac{M_\text{BH}}{M_\odot} = 10^a \left(\frac{\text{FWHM}}{1,000\text{ km s}^{-1}}\right)^b \left(\frac{\lambda L_\lambda}{10^{44}\text{ erg s}^{-1}}\right)^c
\end{equation}
In this work, we use the NIRCam grism spectroscopy to measure the FWHM of their {\hb} emission lines 
and estimate the black hole masses accordingly.
We use the empirical relation suggested by \citet{vp06}, which gives 
$a=6.91, b=2, $ and $c=0.5$.

\subsection{Spectral Fitting} \label{sec:bh:spec}
We run MCMC to fit the spectra of the quasars to measure the profile of the {\hb} lines.
The flux model contains the following components:
(1) a continuum described by a single power law; (2) the iron emission lines described using the template from \citet{park22};
(3) the {\hb} and  {\oiii}$\lambda\lambda4959,5007$ emission lines.
All the emission lines are fitted as two Gaussian components to model the complex broad line profiles seen in high-redshift quasars \citep[e.g.,][]{yang23}.
The parameters of the spectral model include the amplitude and power-law index of the continuum, as well as the fluxes, redshifts, and widths of the emission lines. We adopt flat priors for all these parameters.
We further fix the flux ratio between the {\oiii}4959 and the {\oiii}5007 lines to be $1:3$ \citep{storey00}
and require that the two lines have the same redshifts and widths for both Gaussian components.
The wavelength ranges to be fitted depend on the redshift of the quasars; specifically, we fit the window $3.20 \mu\text{m}<\lambda<3.85 \mu\text{m}$ for J0148+0600, $3.40 \mu\text{m}<\lambda<4.05 \mu\text{m}$ for J1120+0641, and $3.30 \mu\text{m}<\lambda<3.95 \mu\text{m}$ for all the other quasars.
For J1120+0641, the {\hgamma} line is redshifted to the wavelength window, and the {\oiii} lines fall on the edge of the transmission curve. We thus add the {\hgamma} line to the flux model and fix the {\oiii} redshift to ensure a successful fit.

\begin{figure*}
    \centering
    \includegraphics[width=1\linewidth]{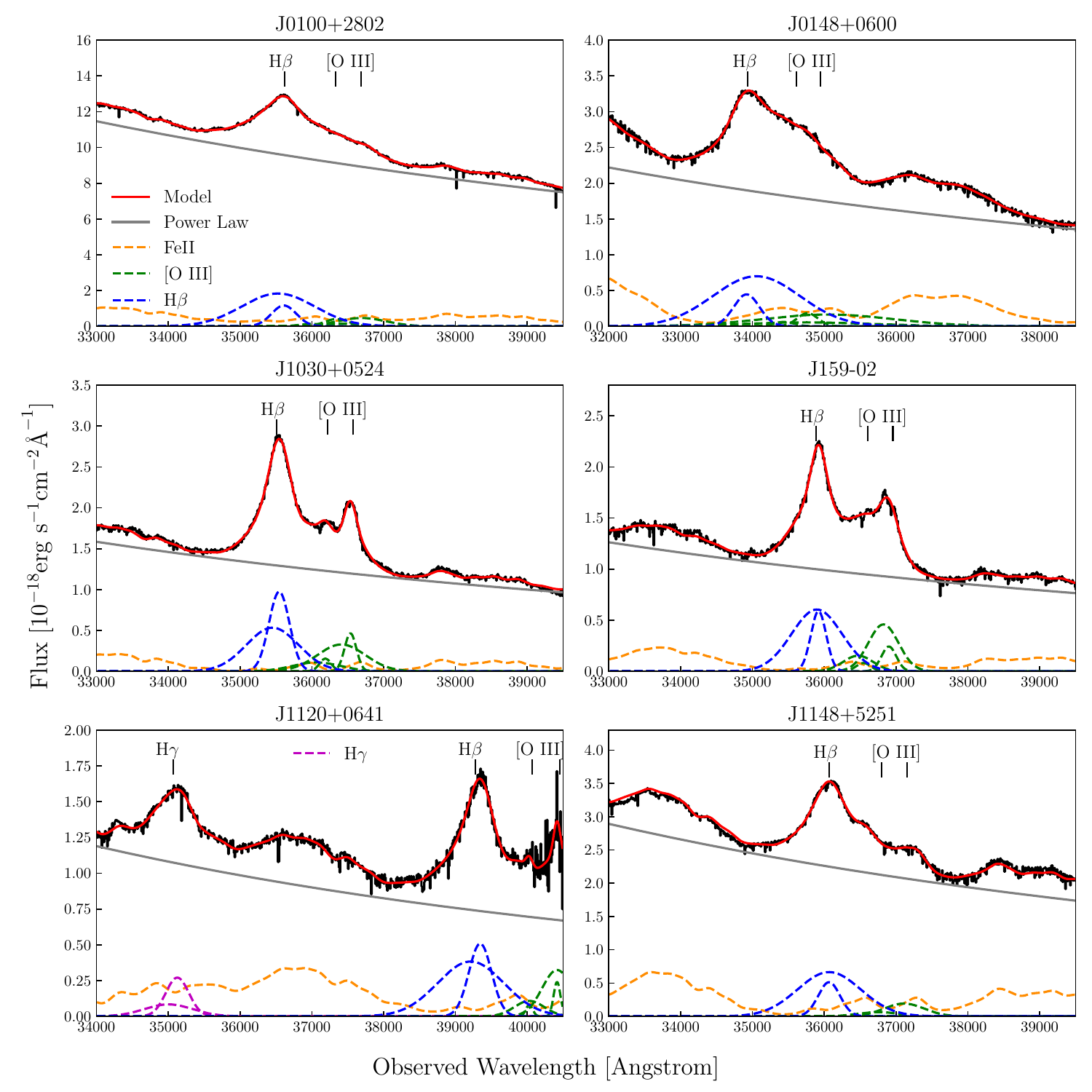}
    \caption{The NIRCam grism spectra of the six quasars in our sample and the best-fit models. We fit the quasar spectra as a continuum power law (gray) plus iron emission lines (orange), the {\hb} lines (blue), the {\oiii} lines (green), and the {\hgamma} lines (magenta) for J1120+0641. We use the widths of the {\hb} lines to measure the mass of the SMBHs.}
    \label{fig:1dspec}
\end{figure*}

Figure \ref{fig:1dspec} shows the spectra and the best-fit flux models of the quasars.
We measure the FWHM of the {\hb} line and the continuum luminosity at rest-frame 5100{\AA} $(L_{5100}=\lambda L_{\lambda}(5100\text{\AA})$,
and calculate the SMBH masses using Equation \ref{eq:mbh}.
We also estimate the bolometric luminosities of the quasars using $L_{5100}$ assuming a bolometric correction of 9.26 \citep{runnoe12},
and compute the Eddington ratios of the quasars.
These results are listed in Table \ref{tbl:properties},
and all the data will be available online after the paper is accepted for publication.

The {\oiii} line profiles have been used to indicate possible outflows driven by the quasars \citep[e.g.,][]{yang23}.
The quasars in the EIGER sample exhibit a variety of {\oiii} line profiles.
J0100+2802, J0148+0600, J1030+0524, and J159-02 exhibit two broad {\oiii} components, while J1148+5251 only has one broad {\oiii} component.
The {\oiii} lines of J1120+0641 are redshifted to the edge of the F356W filter and cannot be well-measured.
This work focuses on the {\hb}-based black hole masses,
and we leave more detailed analysis of the {\oiii} emission lines to future studies.

\subsection{Comparison of Different Black Hole Mass Indicators} \label{sec:bh:comparison}

Previous studies have suggested that {\hb} is a more reliable SMBH mass indicator compared to other broad emission lines \citep[e.g.,][]{shen12}.
However, the {\hb} line of $z\gtrsim6$ objects is not observable with ground-based facilities due to atmospheric absorption,
and previous studies of $z\gtrsim6$ quasars have been using the {\mgii} line to measure the SMBH masses.
The infrared coverage of {\em JWST} makes it possible to probe the rest-frame optical emission lines from high-redshift quasars.
By analyzing the NIRCam grism spectroscopy of eight quasars at $z>6$, 
\citet{yang23} showed that {\mgii}-based SMBH masses are systematically higher than the {\hb}-based SMBH masses.
It is thus important to investigate the differences between the SMBH mass estimates of high-redshift quasars indicated by different emission lines.

\begin{figure}
    \centering
    \includegraphics[width=1\columnwidth, trim={0.2cm 0 0 0}, clip]{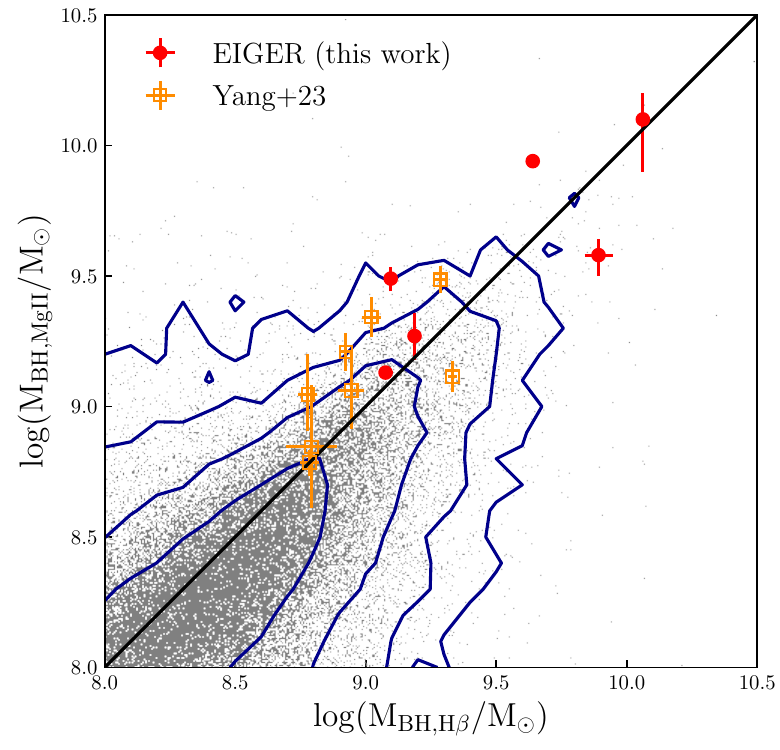}
    \caption{Comparison between the {\mgii}-based and the {\hb}-based SMBH masses. We also include the quasars at $z>6.5$ from \citet{yang23} and the SDSS DR16 quasars \citep{wu22} for comparison. Both the EIGER quasars and the quasars from \citet{yang23}
    have {\mgii}-based BH masses that are larger than their {\hb}-based BH masses. Note that the errorbars represent the random errors from spectral fitting; the systematic uncertainties of BH masses are $\sim0.3$dex.}
    \label{fig:bhcomp}
\end{figure}

Figure \ref{fig:bhcomp} shows the comparison between the {\mgii}-based and the {\hb-}based SMBH mass estimates for the six quasars in this work. We also include $z>6$ quasars from \citet{yang23} and SDSS quasars from \citet{wu22} for comparison. Given the systematic uncertainties of the BH mass estimators $(\sim0.3\text{dex})$, the {\mgii}-based and the {\hb-}based SMBH masses agree with each other. The EIGER quasars have a mean value of
$\log M_{\text{BH, H}\beta} - \log M_{\text{BH, {\mgii}}}=-0.093$; this result is in line with \citet{yang23}, who reported a mean value of $\log M_{\text{BH, H}\beta} - \log M_{\text{BH, {\mgii}}}=-0.13$ for their $z\gtrsim6.5$ quasar sample. These results suggest that the {\mgii-}based BH masses of high-redshift quasars may be systematically larger by $\sim 0.1\text{dex}$ 
than their {\hb-}based BH masses.

\section{The SMBH-Host Galaxy coevolution in the reionization era} \label{sec:coevolve}

\begin{figure*}
    \centering
    \includegraphics[width=.7\linewidth]{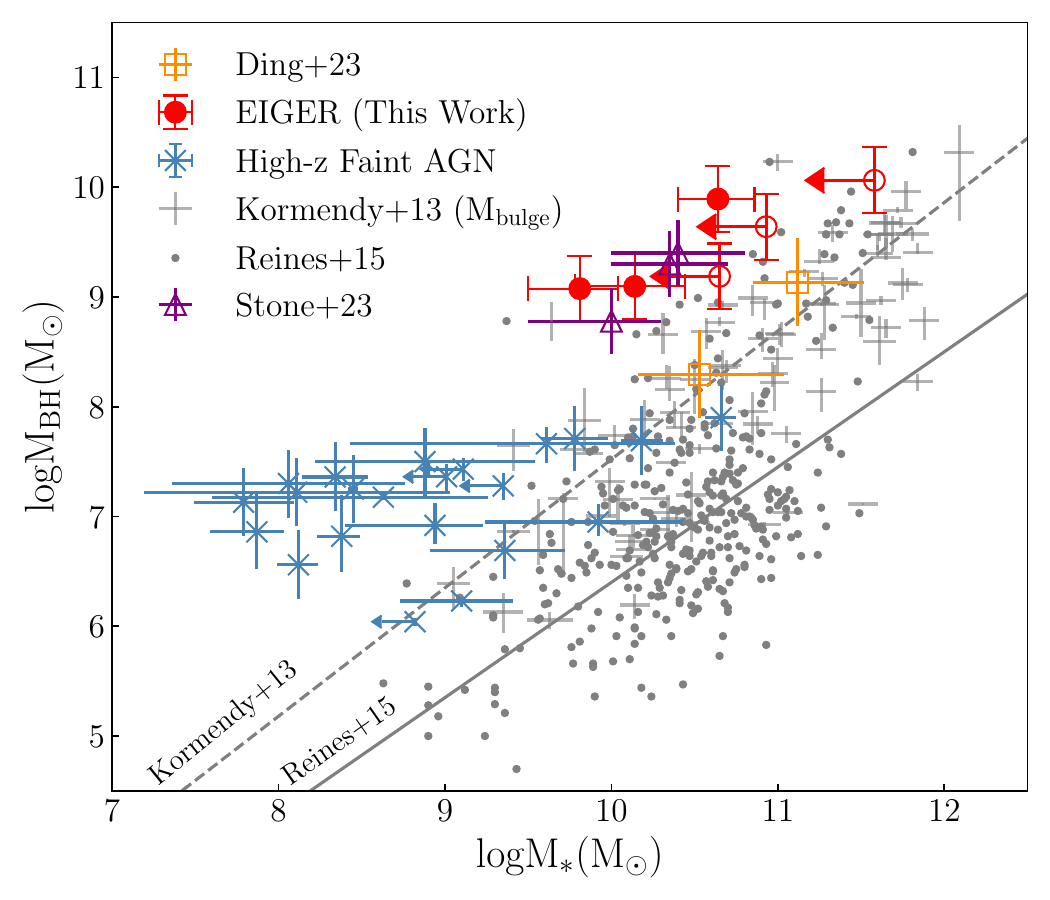}
    \caption{The $M_\text{BH}-M_*$ relation. The filled and open red circles represent EIGER quasars with host galaxy detections and non-detections, respectively. For the EIGER quasars, we use a typical SMBH mass error of $0.3$ dex and use stellar mass errors from the {\texttt{prospector}} models. We include high-redshift quasars from \citet{ding23} and \citet{stone23b}, low-luminosity AGNs from \citet{harikane23} and \citet{maiolino23}, and low-redshift galaxies from \citet{kh13} and \citet{reines15} for comparison. EIGER quasars have $M_\text{BH}/M_*\sim0.15$, which is $\sim2$dex larger than the low-redshift relations. Note that the \citet{kh13} relation uses bulge masses instead of total stellar masses and is located on the left side of the \citet{reines15} relation.}
    \label{fig:msmbh}
\end{figure*}

With the host galaxy stellar masses measured in Section \ref{sec:image} and the black hole masses measured in Section \ref{sec:bh},
we now characterize the position of $z\gtrsim6$ luminous quasars \textcolor{black}{on the $M_\text{BH}-M_*$ plot}.
Figure \ref{fig:msmbh} illustrates the positions of the EIGER quasars on the $M_\text{BH}-M_*$ plot. 
We also include other high-redshift quasar host galaxies and local galaxies from the literature in this Figure\footnote{\textcolor{black}{\citet{stone23b} also measured the stellar masses of J1120+0641 and J1148+5251. Their results are consistent with our measurements or upper limits. For these two objects, we use the $M_\text{BH}$ and $M_*$ evaluated in this work.}}. For the high-redshift quasar sample,
there is a clear trend that more massive galaxies host larger black holes, indicating that the correlation between SMBHs and their host galaxies already exists at $z\gtrsim6$. Meanwhile, the luminous quasars in the EIGER sample have $M_\text{BH}/M_*\sim0.15$, which is $\sim2$ dex higher than the local $M_\text{BH}-M_*$ relation. The quasars with only non-detections of their host galaxies also lie above the local relation according to the upper limits of their stellar masses. This comparison suggests that SMBHs in luminous quasars might have experienced early growth compared to their host galaxies' star formation \citep[e.g.,][]{volonteri12}.


Several previous studies have also used {\em JWST} NIRCam imaging to measure the stellar masses of high-redshift quasars and AGNs. 
\citet{stone23} reported NIRCam observations of a sub-Eddington quasar at $z=6.25$, which also has an overmassive black hole compared to local galaxies.
\citet{ding23} measured the host galaxy fluxes of two low-luminosity quasars at $\sim6.3$, which have small black holes $(M_\text{BH}\sim10^8-10^9M_\odot)$ and massive host galaxies $(M_*\gtrsim10^{10.5}M_\odot)$.
\citet{ding23} argued that the two quasar host galaxies are consistent with the local $M_\text{BH}-M_*$ relation after correcting for the selection effects.

Recent {\em JWST} observations have revealed a population of faint broad line AGNs at $z\gtrsim4$ \citep[also known as the ``little red dots", e.g.,][]{kocevski23,matthee23, maiolino23}.
\citet{harikane23} analyzed a sample of low-luminosity AGNs at $4<z<7$, which have relatively lower $M_\text{BH}/M_*\sim0.01$.
The broad-line AGN at $z=8.679$ reported by \citet{larson23} have $M_\text{BH}/M_*\approx0.3\%$.
However, \citet{furtak23} reported a low-luminosity AGN at $z=7.0451$ with $M_\text{BH}/M_*\gtrsim3\%$.
These results show the large diversity of high-redshift SMBHs and their host galaxies, which might have experienced a variety of growth histories.



Some previous studies have used phenomenological approach to understand the redshift evolution of the $M_\text{BH}-M_*$ relation.
For example, \citet{caplar18} developed an analytical method to derive the $M_\text{BH}-M_*$ relation that matches the observed star formation rate density and quasar luminosity function, who suggested that the BH-to-host mass ratio should be larger at higher redshift, following an evolution of $M_\text{BH}/M_*\propto(1+z)^{2.5}$.
\textcolor{black}{\citet{pacucci23} showed that low-luminosity AGNs at $4<z<7$ recently discovered by {\em JWST} violate the local $M_\text{BH}-M_*$ relation by $\>3\sigma$.}
In contrast, \citet{zhang23} showed that the $M_\text{BH}-M_*$ only evolves mildly at $z<10$ using the empirical model {\textsc{Trinity}}, which was designed to match observables including quasar luminosity functions, quasar probability distribution functions, active black hole mass
functions, local SMBH mass–bulge mass relations, and the SMBH mass distributions of high-redshift bright quasars.
This comparison again shows that it is not a trivial task to correctly characterize the high-redshift $M_\text{BH}-M_*$ relation.


The stellar masses of the high-redshift quasar host galaxies still have large uncertainties.
One of the main reasons is that we only have \textcolor{black}{two or three photometric points} for the SED fitting. By improving the PSF models for image fitting, it should be possible to \textcolor{black}{improve the accuracy of the host galaxy flux measurements and include more photometric points for the SED fitting}. Future observations with NIRSpec IFU will provide emission line fluxes (like the {\oiii} line) of the quasar host galaxies \citep[e.g.,][]{marshall23}, which will put stronger constraints on the stellar masses and other properties of the quasar hosts.


\subsection{Estimating the Selection Bias}

It is worth noticing that selection effects can lead to \textcolor{black}{apparent} overmassive black holes in flux-limited samples of quasars \citep{lauer07}. Specifically, quasars with larger SMBHs are generally more luminous and are more likely to be detected in flux-limited surveys. 
\citet{zhang23} showed that, for luminous AGNs with $L_\text{bol}>10^{46}\text{erg s}^{-1}$ and $M_*\sim10^{10}M_\odot$, this selection effect leads to a SMBH sample that is $\sim1$dex more massive than the true $M_\text{BH}-M_*$ relation. 
However, it is still unclear how to quantify this selection effect and unveil the intrinsic redshift evolution of the $M_\text{BH}-M_*$ relation.
Following the method in \citet{li22}, \citet{ding23} found that the two $z\sim6.4$ quasars in their sample are consistent with no redshift evolution of the $M_\text{BH}-M_*$ relation. In contrast, \citet{stone23} argued that the large $M_\text{BH}/M_*$ values of high-redshift quasars can only be partially explained by selection effects. 

\textcolor{black}{
The EIGER quasars are among the most luminous quasars at $z\gtrsim6$ \citep[e.g.,][]{fan22} and are thus subject to significant selection bias. 
We follow the method described in \citet{li22} to estimate the impact of the selection bias on our results.
To do this, we first generate a sample of mock galaxies by randomly drawing $M_*$ values from the $z\sim6$ stellar mass function suggested by \citet{song16}. We then assign a $M_\text{BH}$ to each mock galaxy following the low-redshift $M_\text{BH}-M_*$ relation in \citet{zhuangho23}:
\begin{equation}
    \log M_\text{BH}=1.49\log(M_*/10^{11}M_\odot)+7.81
\end{equation}
with a scatter of 0.51 dex. We further assume that the Eddington ratios of these SMBHs follow the distribution of $z\sim4.75$ AGNs measured by \citet{kelly13}. We convert the bolometric luminosities of these SMBHs to $L_{5100}$ using the bolometric correction in \citet{runnoe12}. In addition, we add random errors of $0.3$ dex to $M_\text{BH}$ and $0.5$ dex to $M_*$, which mimic the measurement errors of these properties (i.e., the scatters of the single-epoch Virial black hole mass estimator and the systematic error in SED fitting).
}

\begin{figure}
    \centering
    \includegraphics[width=1\linewidth]{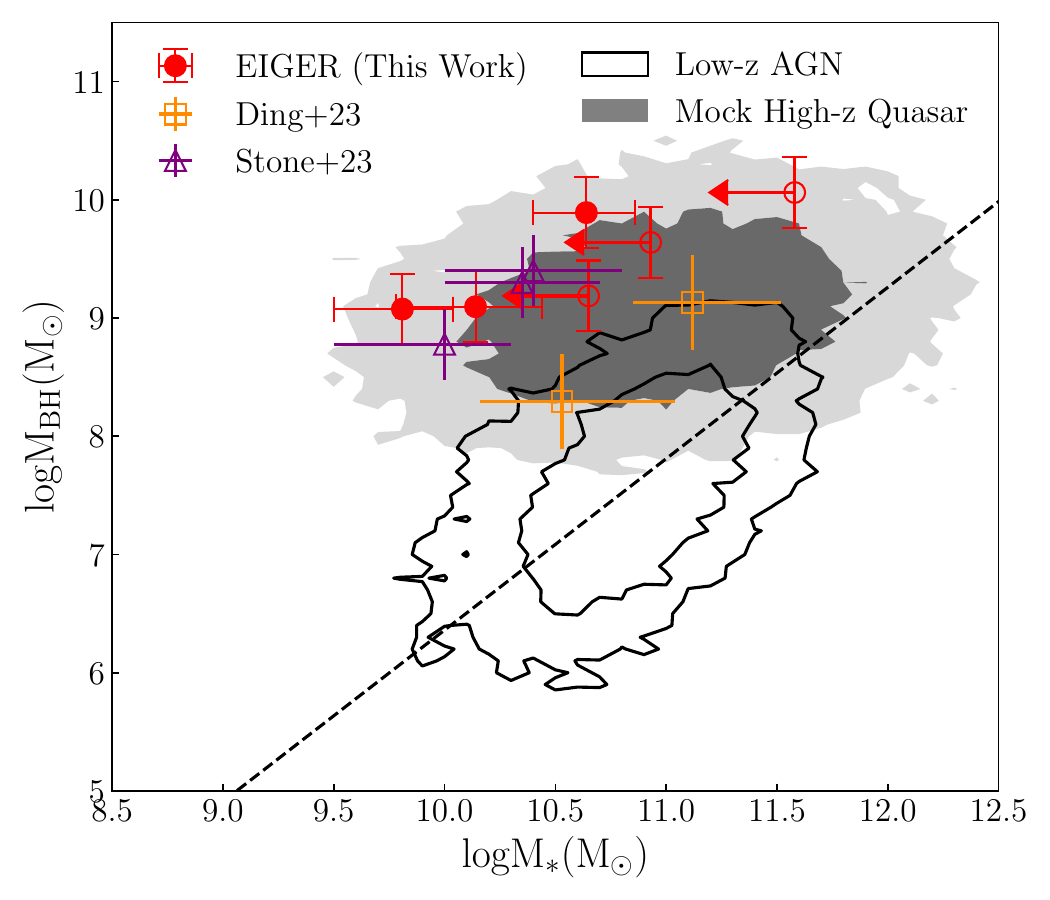}
    \caption{\textcolor{black}{Evaluating the selection bias of the luminous quasar sample. The black contours represent the distribution of low-redshift AGNs in \citet{zhuangho23}. The  filled contours represent the sample of mock quasars at $z\sim6$ with $L_{5100}>10^{46}\text{ erg s}^{-1}$, following the method described in \citet{li22}. From inside to outside, the contours contain 68\% and 95\% of objects (corresponding to $1\sigma$ and $2\sigma$ levels). Due to the selection bias, the mock quasar sample lies above the low-redshift $M_\text{BH}-M_*$ relation. Nevertheless, the quasars in this work still have larger $M_\text{BH}/M_*$ compared to the median of the mock quasar sample.}}
    \label{fig:bias}
\end{figure}

\textcolor{black}{Since the quasars in our sample have $L_{5100}>10^{46}\text{erg s}^{-1}$, we focus on mock galaxies in this luminosity range. The result is shown in Figure \ref{fig:bias}. The solid and filled contours represent low-redshift AGNs and mock high-redshift quasars, respectively. Due to selection bias, high-redshift mock quasars have $M_\text{BH}/M_*$ ratios that are $\sim1$dex higher than the local relation. Individual EIGER quasars locate outside the 68\% confidency contour but fall inside the 95\% confidency contour. Nevertheless, the $M_\text{BH}/M_*$ ratios of the entire EIGER sample are still $\sim1$dex higher than the median of the mock quasars. 
This comparison disfavors the hypothesis that the overmassive black holes seen in high-redshift quasars are purely due to selection effects, and hints at a possible redshift evolution of the $M_\text{BH}/M_*$ ratio. 
This result is consistent with \citet{pacucci23}, who previously inferred a significant redshift evolution of the $M_\text{BH}-M_*$ relation based on the high-redshift faint AGNs discovered by {\em JWST}.}

\textcolor{black}{The high-redshift quasars in \citet{stone23b} have similar $M_\text{BH}/M_*$ ratios to the EIGER quasars, while the two $z\sim6.3$ quasars reported by \citet{ding23} are consistent with the mock quasars. However, we notice that the quasars in \citet{ding23} are fainter $(L_{5100}\sim10^{45}\text{ erg s}^{-1})$ and thus have less severe selection biases compared to the luminous quasars in our sample.}

\textcolor{black}{Finally, we note that the EIGER quasars are not selected to be a flux-limited sample. The mock catalog method in this subsection can only provide a rough estimate of the selection effect. Accurate characterization of the selection bias requires a more uniformly selected sample and a more careful analysis \citep[see, e.g., ][for relevant discussions]{pacucci23, zhang23, zhang23b}, which is beyond the scope of this work. Figure \ref{fig:bias} demonstrates the need for a larger sample of high-redshift quasars with more accurate $M_\text{BH}$ and $M_*$ measurements.}



\section{Discussion}  \label{sec:discuss}

\subsection{Notes on Individual Objects} \label{sec:image:individual}

\subsubsection{J0100+2802} \label{sec:J0100}

J0100+2802 $(z=6.31)$ is a hyperluminous quasar discovered by \citet{wu15}.
Due to its high luminosity, J0100+2802 is saturated in all three bands in EIGER imaging, 
and we are not able to detect its host galaxy.
ALMA observations have suggested that the host galaxy of J0100+2802 
is a starburst galaxy with a star formation rate of $\sim850 M_\odot/\text{yr}$ \citep{wang19b}, which exhibit clumpy structures \citep{fujimoto20}.

\citet{eiger3} reported the NIRCam observations and ground-based spectroscopy of J0100+2802, who showed that the images of this quasar are well-described by a point source, and the {\civ}, {\mgii}, and {\hb} emission lines give consistent estimated BH masses. We note that the data used in this work was produced by a later version of the {\texttt{jwst}} pipeline compared to \citet{eiger3}. The spectral fitting of this work is also slightly different from that of \citet{eiger3}; we fix the line ratios and profiles of the {\oiii} doublet in the spectral fitting, while \citet{eiger3} left all emission line parameters free. 
Consequently, the spectral properties reported in this work are slightly different from that in \citet{eiger3}.

\subsubsection{J0148+0600} \label{sec:J0148}

J0148+0600 $(z=5.977)$ was initially discovered by \citet{jiang15}.
This quasar is known for the large Gunn-Peterson trough in its spectruum \citep[e.g.,][]{becker15}
The PSF-subtracted images of J0148+0600 (Figure \ref{fig:J0148images}) indicate that its host galaxy has a regular elliptical shape. 
J0148+0600 has two projected companions in the north and the south directions, and
it is unclear whether these two objects are associated with the quasar.

\subsubsection{J1030+0524} \label{sec:J1030}

J1030+0524 $(z=6.304)$ was initially reported by \citet{fan01}.
The image fitting process returns a non-detection of the quasar host galaxy according to the criteria in Section \ref{sec:image:fitting}. Nevertheless, the PSF-subtracted images clearly show extended emissions around the quasar, which have consistent shapes in all three bands. These features look like tidal tails and suggest a recent galaxy merging event.
Note that we only got the data from three visits for J1030+0524 by the time of writing this paper.

\subsubsection{J159--02} \label{sec:J159}

J159--02 $(z=6.381)$ was initially reported by \citet{banados16}.
The best-fit image model suggests a regular-shaped host galaxy with large radii.
The PSF-subtracted image shows a projected companion galaxy. 
By the time of writing this paper, J159--02 has only been observed by two visits, where the quasar is located on module A.

\subsubsection{J1120+0641} \label{sec:J1120}

J1120+0641 $(z=7.08)$ was initially reported in \citet{mortlock11}. 
The PSF-subtracted images of J1120+0641 show irregular features, 
suggesting that the quasar might have experienced a recent galaxy merger.
We also notice that the quasar is $\sim0\farcs5$ away \textcolor{black}{(about 3.9 kpc at $z=7.08$)} from the center of the host galaxy. 

\citet{venemans17} reported the high-resolution ALMA observation of J1120+0641.
The sub-mm continuum of J1120+0641 has a deconvolved FWHM of 
$1.24~\text{kpc}\times0.83~\text{kpc}$, which is much smaller than the near-IR emission as revealed by NIRCam imaging. This result indicates that the sub-mm emission and the rest-frame optical emission come from different regions in the quasar host galaxy. 
\citet{venemans17} estimated the dynamical mass $(M_\text{dyn}<4.3\times10^{10}M_\odot)$, 
dust mass $(M_\text{dust}\sim(0.8-4)\times10^{8}M_\odot)$, and gas mass $(M_\text{gas}\lesssim2\times10^{10}M_\odot)$ of the quasar host.
These numbers are in line with the SMBH mass $(M_\text{BH}=1.2\times10^9M_\odot)$ 
and the stellar mass $(M_*=5.6\times10^{9}M_\odot)$ estimated in this work.

We notice that the 
\citet{venemans17} also reported an offset of $\sim0\farcs5$ between the quasar (measured by ground-based imaging) and the host galaxy (measured by ALMA). \textcolor{black}{Specifically, the sub-mm emission locates $0\farcs22$ west and $0\farcs49$ south from the optical quasar. This offset is roughly consistent with our result (see Figure \ref{fig:J1120images}). We note that, due to the limited spatial resolution and astrometric accuracy, \citet{venemans17} were not able to confirm this offset with sufficient significance.}

\citet{bosman23} reported the {\em JWST} Mid-Infrared Instrument (MIRI) spectrum of J1120+0641, which covers observed wavelengths $4.9<\lambda_\text{obs}<27.9~\mu$m. 
\citet{bosman23} reported an H$\alpha$-based BH mass of $1.5\times10^9M_\odot$, 
consistent with the {\hb}-based results within scatters. \citet{bosman23} also found a redshifted H$\alpha$ core component $(\Delta v=-315\pm37\text{ km s}^{-1})$,
which was interpreted as a possible sign of a recoiling black hole by the authors. This scenario is in accordance with the offset between the quasar and the host galaxy emission.

\subsubsection{J1148+5251} \label{sec:J1148}

J1148+5251 $(z=6.42)$ was initially reported by \citet{fan03}.
The image fitting procedure returns a non-detection of the host galaxy.
Nevertheless, the PSF-subtracted image in the F356W band exhibit diffused emission extending from the northeast to the southwest of the quasar. We notice that this emission is absent in the F200W and the F115W images, which suggests that the emission might be dominated by {\hb} or {\oiii} nebular lines. 

\subsection{Systematic Errors and Possible Improvements of the Image Fitting Method}

Measuring the host galaxies of luminous quasars is a challenging task. Given the strong fluxes of the quasars, optimal PSF modeling and image fitting are needed to reveal the emission from the quasar host galaxies.
In this Section, we discuss the PSF modeling and image fitting methods in previous studies and this work, as well as possible improvements to these methods.


We first consider the method to build PSF models. Although recent {\em JWST} observations of quasar host galaxies have been using bright stars to construct PSF models, the exact methods adopted by these studies have some noticeable differences. For example,
\citet{ding23} identified bright stars in the quasar's image, then chose the best five stars  that give the smallest $\chi^2$ in the image fitting as PSFs.
\citet{stone23} explicitly obtained images of a bright star and use the images as PSF, instead of using stars in the quasar's image.
\citet{zhuang23} tested three tools for PSF modeling, including {\texttt{SWarp}} \citep{swarp}, {\texttt{PSFEx}} \citep{psfex}, and {\texttt{photutils}}.
\citet{zhuang23} suggested that {\texttt{PSFEx}} had the best performance,
and that host galaxies with $F_G/F_Q\sim10\%$ can be securely detected with sufficient imaging depth.
 

In this work, we use {\texttt{photutils}} to construct PSF models, where we gather stars from all quasar fields and all visits as PSF stars.
The key feature of this work is that 
we estimate error maps of PSF models and add these errors into image fitting.
We suggest that this step is critical and should be included in similar studies in the future. 
Specifically, if we only consider random noises when fitting the quasar images,
the inaccuracies of the PSF models will have a substantial effect on the fitting result and will bias the estimated host galaxy fluxes.
By including the PSF error maps in image fitting, we give lower weights to pixels that have larger PSF uncertainties and make the resulting host galaxy measurements less biased.

Another systematic error is the SED mismatch between the quasars and the PSF stars. Since quasars and stars have different SEDs in the near-IR, the broad-band PSF of these objects should have different shapes. To investigate this effect, we use {\texttt{webbpsf}} to generate two PSFs corresponding to two types of SEDs: (1) an M dwarf from the stellar spectrum library by \citet{ck03}; (2) a quasar at $z=6$ with the \citet{vdb01} spectral template. For both the F356W and the F200W band, the differences between the two PSFs have similar levels to the PSF error maps described in Section \ref{sec:image:psf}. We thus expect that the impact of SED differences should be comparable to that of PSF error maps. The uncertainties introduced by the PSF error maps can be estimated by the MCMC samples from a {\em single} visit, which introduces an error of $\sigma_{m, \text{host}}\sim0.02$ for the host galaxy magnitude measurements. This value is much smaller than the magnitude errors we report in Table \ref{tbl:imagefitting}, which are dominated by the inter-visit systematic uncertainties.

It is still unclear what is the optimal way to model the PSF of NIRCam. In particular, the PSF of NIRCam depends on the position on the detector, the flux of the source, and the time of the observation. These effects are ignored in this work due to the limited number of PSF stars available, and we leave the detailed analysis of these effects to future studies.

We now consider the method for image fitting. Most studies of quasar host galaxies (including this work) uses a point source plus a S\'ersic profile to describe the image of quasars. The specific setting for the S\'ersic profiles varies between the studies. \citet{ding23} left all S\'ersic parameters free, \citet{zhuang23} fixed the center of the host galaxy to the quasar's position, and this work fix the S\'ersic index of the host galaxy to be $n=1$ (i.e., an exponential disk). Studies based on ground-based imaging have used one-dimensional profile fitting, which is a powerful tool for low-redshift quasar host galaxies \citep[e.g.,][]{mat14,yue18}.


The result of this work demonstrates the limitation of the existing image fitting method. Several quasars in the EIGER sample show irregular emissions in their PSF-subtracted images. For these quasars, a regular S\'ersic profile might not give the correct description of the host galaxy. It is thus desirable to develop image fitting methods that can describe irregular galaxy shapes. Additionally, we suggest that the position of the host galaxy should be left free instead of fixed to the quasar's position since some quasars show large offsets from their host galaxies (like J1120+0641).


Using the {\em bluetides} simulation \citep{bluetides}, 
\citet{marshall21} estimated that the detection limit of high-redshift quasar host galaxies is $F_G/F_Q\sim1\%$ for NIRCam (assuming an exposure time of $10$ ks). This work successfully detects quasar host galaxies that have $F_G/F_Q\sim1\%$ with exposure times of $\sim1.6$ ks for the F356W imaging. With improved PSF models in the future, it is possible to achieve detection limits of $F_G/F_Q\lesssim1\%$, allowing us to measure quasars with fainter host galaxies.



\section{Conclusion}  \label{sec:sum}

We present NIRCam observations of six quasars at $z\gtrsim6$ observed by the EIGER project.
We use NIRCam imaging to measure the host galaxy emissions of the quasars, where we fit the quasar images as a point source plus an exponential disk.
We construct PSF models and their error maps using bright stars in the images, and run MCMC to perform image fitting to estimate the fluxes of the quasar host galaxies.
We use NIRCam grism spectra to measure the profile of the broad {\hb} emission lines and calculate the SMBH masses of the quasars.
The main results of this work are as follows:

\begin{enumerate}
    \item \textcolor{black}{We detect the host galaxy of J0148+0600 in all the three bands, as well as the host galaxies of J159--02 and J1120+0641 in the F200W and the F356W bands. We report tentative detections for the host galaxies of J159--02 and J1120+0641 in the F115W band.} These quasars have host-to-quasar flux ratios of $\sim1\%-5\%$. SED fitting shows that these quasar host galaxies have $M_*\gtrsim10^{10}M_\odot$. 
    \item We report non-detections for the host galaxies of J0100+2802, J1030+0524, and J1148+5251. We also estimate the upper limits of the fluxes and stellar masses of their host galaxies. The PSF-subtracted image of J1030+0524 and J1148+5251 show diffused emissions around the quasar, which might come from the tidal tails of the ongoing galaxy merger or extended {\oiii} emissions around the quasar.
    \item We compute the black hole masses of the six quasars using their {\hb} emission lines. The {\hb}-based black hole masses of these quasars are slightly smaller than the {\mgii}-based ones by $\sim 0.1\rm dex$, which is consistent with the results by \citet{yang23}.
    \item The quasars in the EIGER sample have $M_\text{BH}/M_*\sim0.15$, which is \textcolor{black}{$\sim2$ dex} larger than low-redshift galaxies with similar stellar masses. This comparison suggests that high-redshift quasars might have experienced early SMBH growth compared to the star formation of their host galaxies. \textcolor{black}{Selection bias also contributes to the high $M_\text{BH}/M_*$ ratios of luminous quasars. However, the $M_\text{BH}/M_*$ ratios of EIGER quasars are larger than the mock luminous quasar sample even after the selection bias is considered, hinting at a possible redshift evolution of the $M_\text{BH}-M_*$ relation.} 
\end{enumerate}

The EIGER quasars are among the most luminous quasars at $z\gtrsim6$. The detection of their host galaxies illustrates the promising potential for {\em JWST} to build a large sample of high-redshift quasars with host galaxy detections in the near-IR. The PSF models of NIRCam will be improved by future observations and analyses, which will enable  more accurate characterization of the high-redshift $M_\text{BH}-M_*$ relation. Meanwhile, {\em JWST} observations have discovered several AGNs at even higher redshifts \citep[e.g.,][]{furtak23,goulding23,larson23,kokorev23}. Follow-up analysis of these AGNs will characterize the co-evolution of SMBHs and their host galaxies at even earlier cosmic times, approaching the origin of the $M_\text{BH}-M_*$ relation that already has its shape at $z\sim6$.

\begin{acknowledgments}
\textcolor{black}{We thank the referee for the valuable comments on this paper.}
We thank John Silverman, Madeline Marshall, Ming-Yang Zhuang, Weizhe Liu and Jinyi Yang for inspiring discussions and suggestions. DK thanks the support from JSPS KAKENHI Grant Number JP21K13956.
This work is based on observations made with the NASA/ESA/CSA James Webb Space Telescope. The data were obtained from the Mikulski Archive for Space Telescopes at the Space Telescope Science Institute, which is operated by the Association of Universities for Research in Astronomy, Inc., under NASA contract NAS 5-03127 for JWST. These observations are associated with program ID $\#1243$. 
\end{acknowledgments}

%

\vspace{5mm}
\facilities{JWST(NIRCam)}


\software{astropy \citep{2013A&A...558A..33A,2018AJ....156..123A},  
            psfMC\citep{psfmc},
            webbpsf\citep{webbpsf}, jwst
          }



\appendix

\section{Validating the Quasar Host Galaxy Detections}

\label{sec:image:fitting:limit}


\textcolor{black}{Given the small flux ratios between the detected host galaxies and the quasars,
we need to ensure that the detected quasar host galaxy emissions are real and are not results of PSF model inaccuracies.
We perform this test by running the image fitting procedure on the PSF stars selected in Section \ref{sec:image:psf}. 
For each PSF star, we first build a new empirical PSF model
by excluding the star from the PSF star list, 
then fit the star as a point source plus an exponential disk and a sky background.
If the PSF models are accurate, we will get non-detections for the exponential disk components.
In reality, we may get positive fluxes for the exponential disk components
due to the inaccuracies of the PSF models, according to which
we can examine whether the quasar host galaxy detections we report are reliable.}

\begin{figure*}
    \centering
    \includegraphics[width=0.32\linewidth]{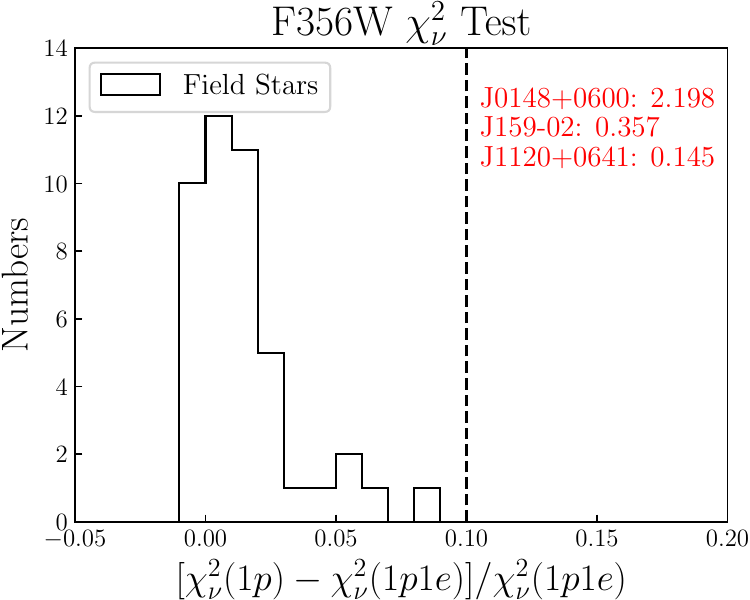}
    \includegraphics[width=0.32\linewidth]{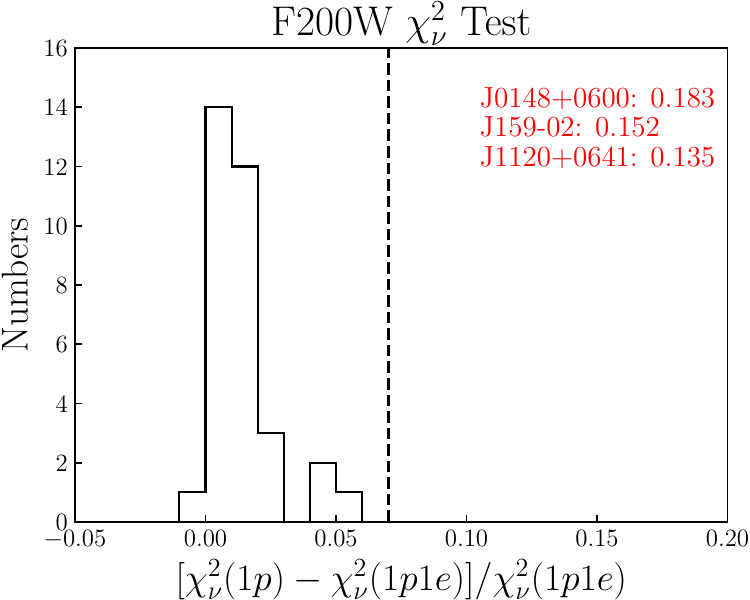}
    \includegraphics[width=0.32\linewidth]{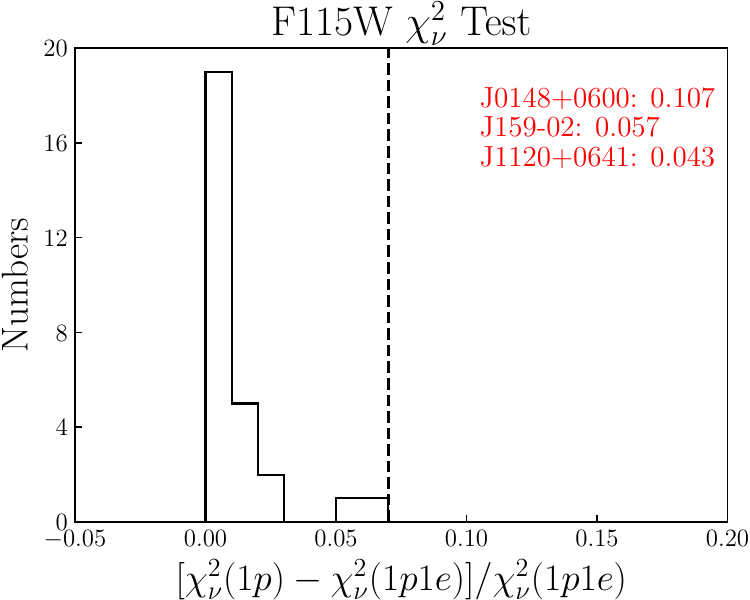}
    \caption{Testing the reduced $\chi^2$ of the quasars and stars. The histograms present the distribution of $\Delta \chi^2_\nu/\chi^2_\nu$ (Equation \ref{eq:rchisq}), and the black dashed line marks the criterion for successful host galaxy detection. The $\Delta \chi^2_\nu/\chi^2_\nu$ of the quasars with host galaxy detection, averaged over all visits, are listed in red text. In the F356W and the F200W bands, all three quasars have $\Delta \chi^2_\nu$ that are much higher than the PSF stars. In the F115W band, only J0148+0600 passes the detection criteria in the F115W band, and J159--02 and J1120+0641 have $\Delta \chi^2_\nu$ that are similar to some stars.}
    \label{fig:chi2test}
\end{figure*}

\textcolor{black}{We first examine the reduced $\chi^2$ values for stars and quasars. 
The result is shown in Figure \ref{fig:chi2test}.
Quasars have significantly higher $\Delta\chi^2_\nu/\chi^2_\nu$ values than stars. Specifically,
all stars have $\Delta\chi^2_\nu/\chi^2_\nu<0.1$ in the F356W band and $\Delta\chi^2_\nu/\chi^2_\nu<0.07$ in the F200W and the F115W band; in contrast,
the three quasars with host galaxy detections have $\chi^2_\nu/\chi^2_\nu>0.1$ in the F356W and F200W bands. 
The differences in $\Delta\chi^2_\nu/\chi^2_\nu$ between stars and quasars are smaller in the F115W band. 
Figure \ref{fig:chi2test} suggests that the host galaxy detections for the three quasars are real at least in the F356W and the F200W bands.
These results also explain why we adopt the $\Delta\chi^2_\nu/\chi^2_\nu$ criteria in Section \ref{sec:image:fitting}.}

 {Another useful indicator of the quasar host galaxy detections is the signal-to-noise ratio $(S/N)$. For each quasar, we evaluate the $S/N$ of the host galaxy signal within an ellipse aperture that corresponds to two times the half-light radii. The dashed ellipses in Figure \ref{fig:J0148images} to \ref{fig:J159images} mark these apertures. We notice that, due to the contribution of the PSF errors, the noises at central pixels are much larger than the noises at larger radii and can even be larger than the total flux of the host galaxy. Meanwhile, the ``(Image-PSF)/Error'' columns in Figure \ref{fig:J0148images} to \ref{fig:J159images} show that the host galaxy signal is significant at larger radii where the PSF errors are smaller. Based on this consideration, when evaluating the S/N of quasar host galaxies, we exclude the central regions with $r<2\times\text{FWHM}_\text{PSF}$ where the PSF errors are high. The dashed circles in Figure \ref{fig:J0148images} to \ref{fig:J159images} represent the masked central regions. We then compute the signal $(S)$ and noise $(N)$ in the aperture:}
{
\begin{equation}\label{eq:s/n}
    S = \sum\limits_{(x,y)}{I(x,y)-P(x,y)},
    N = \sqrt{{\sum\limits_{(x,y)}\epsilon_\text{all}(x,y)^2}}
\end{equation}
where $I(x,y)$ is the original image, $P(x,y)$ is the best-fit PSF component, and $\epsilon_\text{all}(x,y)$ is the composite noise map (Equation \ref{eq:adderr}). The $S/N$ of these quasar host galaxies are presented in Figure \ref{fig:J0148images} to Figure \ref{fig:J159images}; these quasar host galaxy detections have $S/N>10$, which again confirms that the host galaxy signals cannot be explained by PSF errors.
}

\begin{figure*}
    \centering
    \includegraphics[width=1\linewidth,trim={0.5cm 0 0.2cm 0},clip]{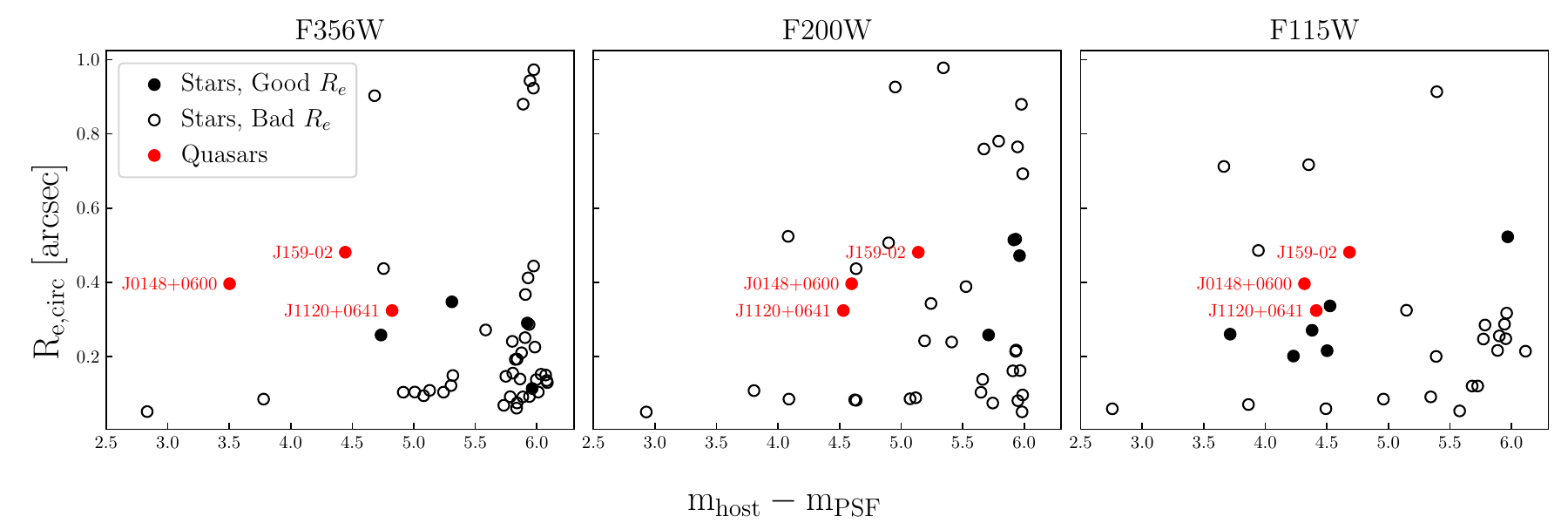}
    \caption{Testing the host galaxy detection reliability using field stars. We fit the stars as a PSF plus an exponential disk and a background component, following the same method we used for quasar image fitting. Open circles mark stars that do not satisfy the radii-related criteria, 
    filled black circles mark stars that pass the radii-related detection criteria,
    and the red circles represent the quasars with host galaxy detections. \textcolor{black}{Most stars in the F356W and the F200W band fail to pass the radii-related criteria, while some stars in the F115W band pass the radii-related detection criteria and have $m_\text{host}-m_\text{PSF}$ similar to the real quasars. This Figure demonstrates the reason why we adopt the radii-related detection criteria for the quasar host galaxies.}}
    \label{fig:psftest}
\end{figure*}

\textcolor{black}{We further examine the best-fit parameters of the exponential disk component for the stars.
The result of the experiment is shown in Figure \ref{fig:psftest}.
 Again, we note that the three bands have distinct PSF star samples due to the magnitude requirements (Section \ref{sec:image:psf});
 in particular, there is no overlap between the PSF stars for the F356W band and the F200W band.
The open black circles mark stars where the best-fit exponential profile does not satisfy the radii-related criteria (criteria (2) and (3) in Section \ref{sec:image:fitting}), for which our image fitting method would report non-detections of the host galaxies.
The solid circles mark stars that satisfy the radii-related criteria.
Note that we do not apply criterion (1) to the stars because most stars were only observed by one visit, 
meaning that we cannot estimate the inter-visit systematic uncertainties for the stars.}

\textcolor{black}{
For the majority of stars, the best-fit exponential profiles have either a very large radii (mimicking a sky background), a very small radii (mimicking a point source), or a small axis ratio (mimicking the PSF residual, especially on the spikes). As a result,
most $(>90\%)$ PSF stars in the F356W band and the F200W band fail to pass criteria (2) and (3) and thus return non-detections of the host galaxy component.
This result explains why we adopt these criteria in Section \ref{sec:image:fitting}.
We also notice that stars in the F356W and F200W bands that pass criteria (2) and (3) have faint best-fit exponential profiles ($\Delta m=m_\text{exp}-m_\text{PSF}\gtrsim5$, where $m_\text{exp}$ stands for the best-fit magnitude of the exponential disk component).
Meanwhile, some stars in the F115W band return an exponential disk component that is $\sim3.5-4.5$ magnitudes fainter than the point source component.
By visually inspecting their PSF-subtracted images, we find that these stars have strong PSF wings and spikes, which mimic the extended exponential disk. This result is also consistent with the finding in Section \ref{sec:image:psf} and in \citet{zhuang23}, i.e., the PSF spatial variations are stronger in bluer bands. }



\textcolor{black}{The red circles in Figure \ref{fig:psftest} represent the quasar host galaxies detected in Section \ref{sec:image:fitting}. 
These quasars pass criteria (2) and (3) in the F356W band, indicating that the host galaxy detections are not results of PSF inaccuracies.
Another piece of evidence for the host galaxy detections is the consistent results from individual visits. Figure \ref{fig:J0148visits} presents the images of J0148+0600 as an example. For the other two quasars, we also find that the PSF-subtracted images from individual visits show similar patterns and that the best-fit host galaxy magnitudes from different visits are consistent within $\lesssim0.3$ magnitudes.}

\textcolor{black}{
We notice that a few stars in the F115W band return $\Delta m$ and $R_e$ values similar to the quasar host galaxies, meaning that 
it is possible that PSF inaccuracies contribute significantly to the F115W fluxes of the quasar host galaxies.
This result is consistent with our finding in Section \ref{sec:image:fitting} that the PSF errors are larger in the F115W band.
Furthermore, J1120+0641 and J159-02 fail to pass the $\Delta \chi^2_\nu$ criteria in the F115W band. We thus  exclude the F115W magnitudes in the SED fitting for these two quasar host galaxies.}


\textcolor{black}{
The next question is whether the reported quasar host galaxy properties (i.e., magnitudes and radii) and their errors are reliable. 
We answer this question using mock observations.
For each quasar, we generate a set of mock images by adding an exponential profile to the PSF star images.
We scale the images and the error maps of the PSF stars to match the total flux of the quasar, and use the best-fit parameters of the quasar host galaxy to generate the exponential profile. We notice that the errors of image fitting can be large for stars that are much fainter than the quasar. To mitigate this effect, we only use stars that have $|m_\text{star}-m_\text{quasar}|<1$ to generate mock observations. The only exception is J0148+0600 in the F200W and F115W bands, where we adopt $-1<m_\text{star}-m_\text{quasar}<2$ to include sufficient numbers of stars.}

\textcolor{black}{We then run image fitting for the mock images to examine whether we can recover the input host galaxy parameters. Following the method used for the quasars,
we leave the exponential disk morphology free when fitting the F356W images, and keep the morphology fixed to the true value when fitting the F200W and the F115W images. We mimic the multiple visits to the quasars by randomly drawing mock observations. Specifically, for one realization, we randomly select $N$ mock observations (where $N$ is the number of visits), then compute the median and the standard deviation of the host galaxy parameters among these observations as the best-fit values and their errors. We generate 100 realizations for each quasar and each band.}

\begin{figure*}
    \centering
    \includegraphics[width=0.32\textwidth]{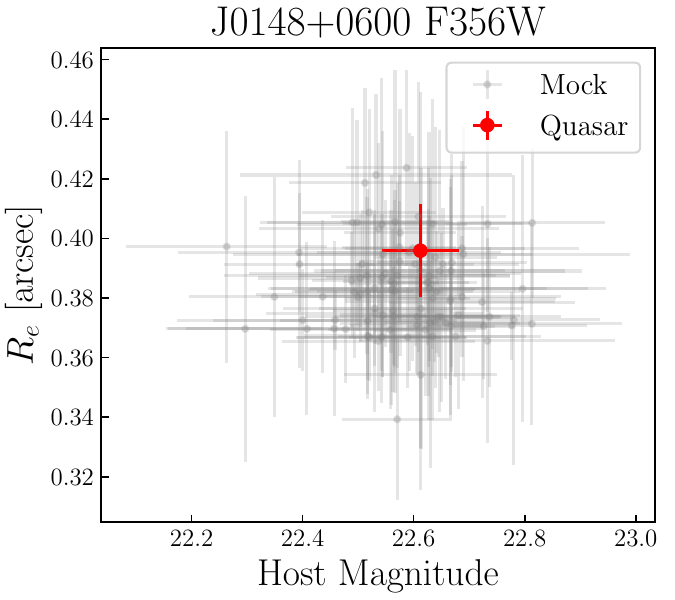}
    \includegraphics[width=0.32\textwidth]{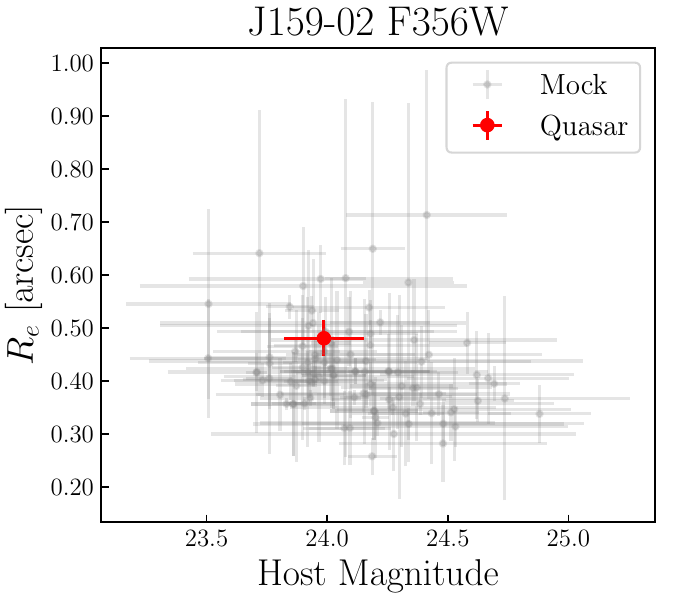}
    \includegraphics[width=0.32\textwidth]{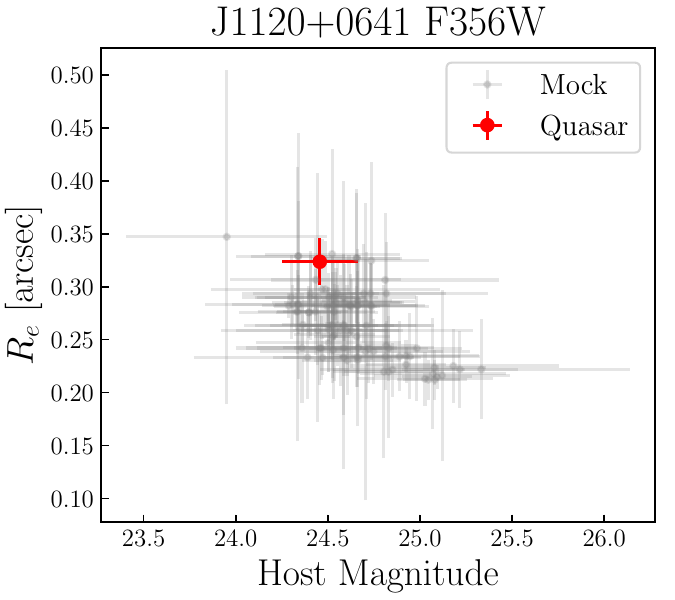}
    \includegraphics[width=0.32\textwidth]{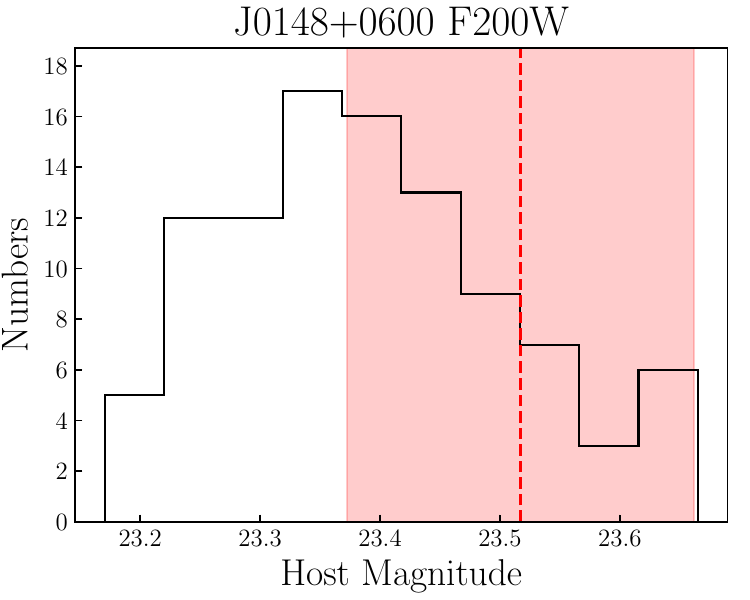}
    \includegraphics[width=0.32\textwidth]{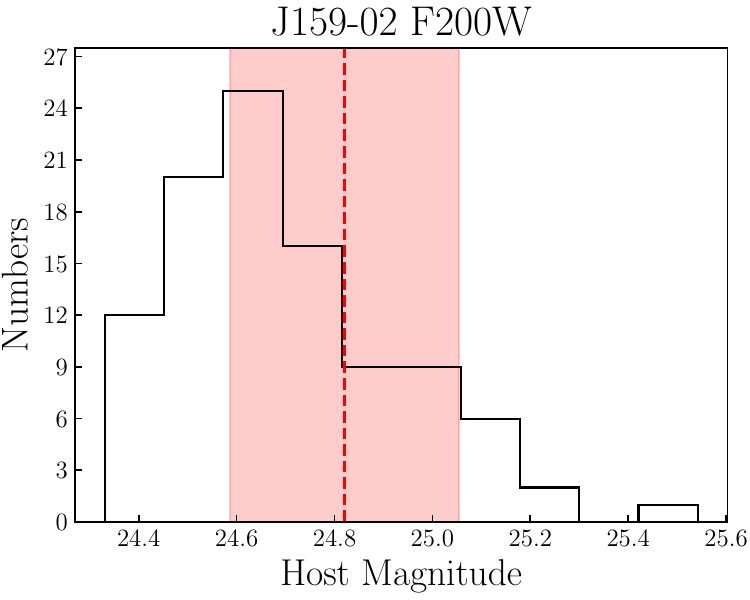}
    \includegraphics[width=0.32\textwidth]{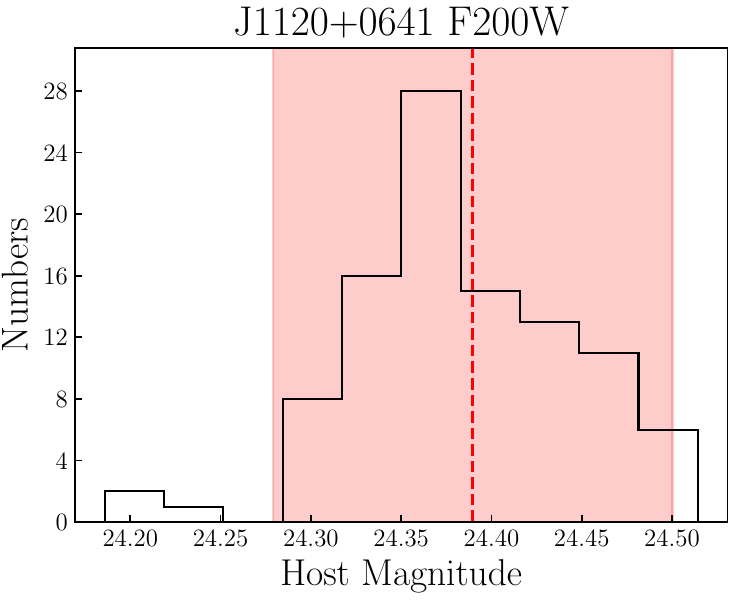}
    \includegraphics[width=0.32\textwidth]{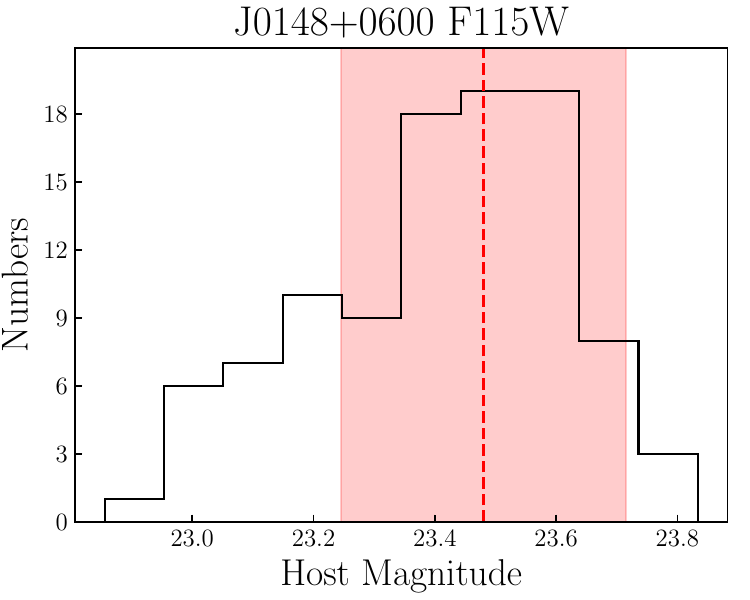}
    \includegraphics[width=0.32\textwidth]{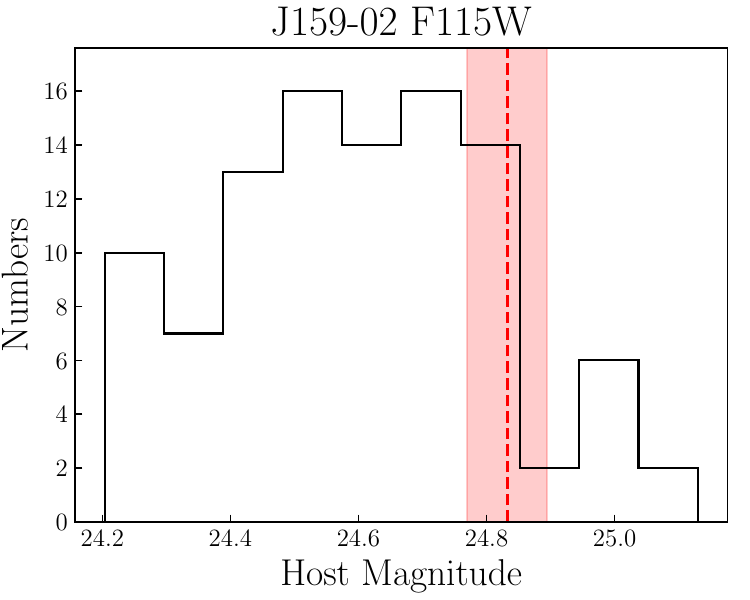}
    \includegraphics[width=0.32\textwidth]{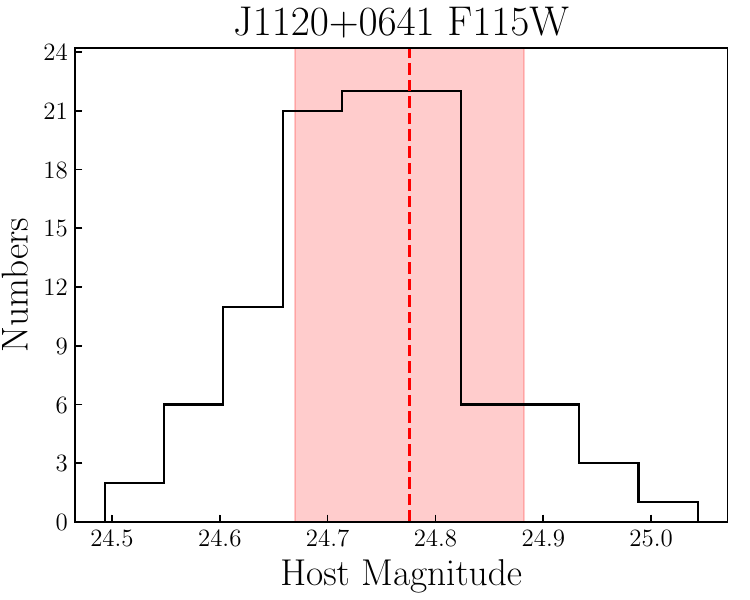}
    \caption{Testing the reliability of the reported quasar host galaxies and radii. See text for details about mock observation generation. {\em Top panel:} the F356W band test, where the morphology and positions of the exponential profile are left free. Each black dot represents one realization, and the red dot represents the real observations on the quasar (which is also the input value to generate mock observations). {\em Middle and Bottom}: the F200W and the F115W band test, where the morphology and positions of the exponential profile are fixed to the input values. The histograms represent the realizations of mock observations, and the red line and the shaded area represent the best-fit host galaxy magnitude and its error from real observations on quasars. The mock observations give exponential profile magnitudes that are consistent with the input value (except for J159--02 in the F115W band, which we report as a tentative detection). This test validates the quasar host galaxy magnitudes and their errors we report in this work.}
    \label{fig:mockobs}
\end{figure*}

\textcolor{black}{Figure \ref{fig:mockobs} presents the results of this test. In the top panel, we examine the best-fit exponential profile magnitudes and radii from mock F356W observations. Each gray point represents one realization, and the red point represents the real observation of the quasar (which is also the input parameter to generate mock observations). The best-fit exponential profile magnitudes of the mock observations are consistent with the input values within the errors, and the radii of the exponential profiles can be recovered with an accuracy of $\lesssim50\%$.
In the bottom two panels, we show the best-fit exponential profile magnitudes for the mock observations in the F200W and the F115W bands. Note that the morphology of the exponential profile is fixed when fitting these two bands. In most cases, the best-fit exponential profile magnitudes from mock observations are consistent with the input value within errors, except for J159--02 in the F115W band, which we report as a tentative detection. 
These results indicate that the host galaxy magnitudes and their errors reported in this work are reliable.}

\textcolor{black}{To summarize, the host galaxy detections of J0148+0600, J159--02, and J1120+0641 are reliable and cannot be explained by PSF inaccuracies. We report host galaxy detections for J0148+0600 in all three bands. For J159--02 and J1120+0641, we report host galaxy detections in the F356W and the F200W bands, as well as tentative detections in the F115W band. Mock observations show that the reported host galaxy magnitudes and their errors are accurate, validating our image fitting method.}



\bibliography{sample631}{}
\bibliographystyle{aasjournal}



\end{document}